\documentclass[twocolumn]{aastex631}
\usepackage{threeparttable}
\usepackage{booktabs}
\usepackage{subfigure}
\usepackage{amsmath}
\usepackage{epstopdf}
\usepackage{color}
\usepackage{enumitem}
\usepackage{tabularx}   
\usepackage{multirow}
\usepackage{makecell}

\usepackage{comment}
\setlist[enumerate]{label=\textbf{\arabic*.}}

\newcommand{\reffig}[1]{Figure \ref{#1}}
\newcommand{\reftable}[1]{Table \ref{#1}}
\newcommand{\refsection}[1]{Section \ref{#1}}
\newcommand{\refsubsection}[1]{Section \ref{#1}}

\newcommand{\refformula}[1]{Equation (\ref{#1})}

\begin{document}
	\title{PISP: Projected-Space Inference of Stellar Parameters}
	\author[0009-0001-9085-8718]{Jun-Chao Liang}
	\affil{National Astronomical Observatories, Chinese Academy of Sciences, Beijing 100101, People’s Republic of China}
	\affil{School of Astronomy and Space Science, University of Chinese Academy of Sciences, Beijing 100049, People’s Republic of China}
	
	\author[0000-0001-7607-2666]{Yin-Bi Li $^\star$}
	\affil{National Astronomical Observatories, Chinese Academy of Sciences, Beijing 100101, People’s Republic of China}
	\email{$^\star$ ybli@bao.ac.cn}
	
	\author[0000-0001-7865-2648]{A-Li Luo $^\star$}
	\affil{National Astronomical Observatories, Chinese Academy of Sciences, Beijing 100101, People’s Republic of China}
	\affil{School of Astronomy and Space Science, University of Chinese Academy of Sciences, Beijing 100049, People’s Republic of China}
	\affil{University of Chinese Academy of Sciences, Nanjing 211135, People’s Republic of China}
	\email{$^\star$ lal@nao.cas.cn}
	
	\author[0000-0002-8913-3605]{Shuo Li}
	\affil{National Astronomical Observatories, Chinese Academy of Sciences, Beijing 100101, People’s Republic of China}
	\affil{School of Astronomy and Space Science, University of Chinese Academy of Sciences, Beijing 100049, People’s Republic of China}
	
	\author[0000-0002-9279-2783]{Xiao-Xiao Ma}
	\affil{National Astronomical Observatories, Chinese Academy of Sciences, Beijing 100101, People’s Republic of China}
	\affil{School of Astronomy and Space Science, University of Chinese Academy of Sciences, Beijing 100049, People’s Republic of China}
	
	\author[0000-0002-8252-8743]{Hai-Ling Lu}
	\affil{National Astronomical Observatories, Chinese Academy of Sciences, Beijing 100101, People’s Republic of China}
	\affil{School of Astronomy and Space Science, University of Chinese Academy of Sciences, Beijing 100049, People’s Republic of China}
	
	\author[0000-0001-5066-5682]{Shu-Guo Ma}
	\affil{National Astronomical Observatories, Chinese Academy of Sciences, Beijing 100101, People’s Republic of China}
	
	\author[0009-0001-0257-3377]{Ming-Hui Jia}
	\affil{Institute of Automation, Chinese Academy of Sciences, Beijing 100190, People’s Republic of China}
	\affil{School of Advanced Interdisciplinary Sciences, University of Chinese Academy of Sciences, Beijing 100049, People’s Republic of China}
	
	\author[0009-0005-6200-2778]{Shuo Ye}
	\affil{National Astronomical Observatories, Chinese Academy of Sciences, Beijing 100101, People’s Republic of China}
	\affil{School of Astronomy and Space Science, University of Chinese Academy of Sciences, Beijing 100049, People’s Republic of China}
	
	\author{Hao Zeng}
	\affil{National Astronomical Observatories, Chinese Academy of Sciences, Beijing 100101, People’s Republic of China}
	\affil{School of Astronomy and Space Science, University of Chinese Academy of Sciences, Beijing 100049, People’s Republic of China}
	
	\author[0009-0009-0513-7042]{Ke-Fei Wu}
	\affil{National Astronomical Observatories, Chinese Academy of Sciences, Beijing 100101, People’s Republic of China}

	\author[0009-0003-6636-8203]{Zhi-Hua Zhong}
	\affil{Center for Spatial Information Science, The University of Tokyo, Tokyo 153-8505, Japan}
	
	\author[0000-0001-8011-8401]{Xiao Kong}
	\affil{National Astronomical Observatories, Chinese Academy of Sciences, Beijing 100101, People’s Republic of China}
	
	\author[0000-0001-7783-9662]{Li-Li Wang}
	\affil{School of Computer and Information, Dezhou University, Dezhou 253023, People’s Republic of China}
	
	\author[0000-0003-0433-3665]{Hugh R. A. Jones}
	\affiliation{School of Physics, Astronomy and Mathematics, University of Hertfordshire, United Kingdom}
	
	\begin{abstract}
			{To improve the accuracy and efficiency of high-dimensional stellar parameter inference in large spectroscopic datasets, we propose a projection-assisted parameter-inference framework---Projected-Space Inference of Stellar Parameters (PISP). PISP constructs an orthonormal basis and optimizes in the projected space, reducing the impact of parameter correlations on inference. The basis is constructed using either principal component analysis (PCA) or the active-subspace (AS) method and is combined with two inference strategies---Non-L1, which optimizes the projection coefficients for a user-specified projected dimensionality, and L1, which introduces L1 regularization in the full projected space to adaptively select projection directions---yielding four strategies: PCA-Non-L1, AS-Non-L1, PCA-L1, and AS-L1. For different computational scenarios, we implement two versions: PISP-CurveFit for fast single-spectrum inference and PISP-Adam for large-scale GPU-parallel inference. Using a fully connected neural network and a residual network as spectral emulators, we evaluate PISP on Kurucz synthetic spectra and on $722{,}896$ APOGEE DR$17$ observed spectra. Compared to the baseline strategy, PISP improves inference accuracy for multiple parameters across all emulator-optimizer combinations. In synthetic data, PCA-L1 performs best, reducing the standard deviation of differences ($\sigma(\Delta)$) by at least $0.01$ dex for $12$ of $20$ elemental abundances, with [N/H], [O/H], [Na/H], [Co/H], [P/H], [V/H], [Cu/H] showing $0.05$--$0.72$ dex reductions. In observed data, PCA-Non-L1 reduces $\sigma(\Delta)$ by $>30$ K for effective temperature and by at least $0.01$ dex for $9$ of $17$ elemental abundances, with [O/H], [Na/H], [V/H] showing $0.05$--$0.20$ dex reductions, while achieving a $\sim$$4\times$ efficiency gain, slightly outperforming PCA-L1.}
	\end{abstract}
	
	
	\section{Introduction}
	In recent years, large spectroscopic surveys such as the Radial Velocity Experiment \citep{Steinmetz2006}, the Sloan Digital Sky Survey (SDSS; \citeauthor{Yanny2009} \citeyear{Yanny2009}; \citeauthor{Majewski2017} \citeyear{Majewski2017}; \citeauthor{Almeida2023} \citeyear{Almeida2023}), the Large sky Area Multi-Object fiber Spectroscopic Telescope (LAMOST; \citeauthor{Zhao2012} \citeyear{Zhao2012}; \citeauthor{Cui2012} \citeyear{Cui2012}; \citeauthor{Luo2015} \citeyear{Luo2015}), the Galactic Archaeology with HERMES (GALAH; \citeauthor{DeSilva2015} \citeyear{DeSilva2015}), the Gaia Radial Velocity Spectrometer (RVS; \citeauthor{GaiaCollaboration2023}\citeyear{GaiaCollaboration2023}; \citeauthor{RecioBlanco2023}\citeyear{RecioBlanco2023}), and the Dark Energy Spectroscopic Instrument (DESI; \citeauthor{DESICollaboration2025} \citeyear{DESICollaboration2025}), among others, have provided an unprecedented observational foundation for studies of stellar physics and Galactic structure.
	
	These datasets enable the inference of key stellar parameters, such as effective temperature ($T_{\mathrm{eff}}$), surface gravity ($\log g$), and the abundances of dozens of elements. These stellar parameters constitute fundamental data of significant importance for many areas of contemporary astronomy and astrophysics, ranging from stellar evolution and the formation of planetary systems to the evolution of the Milky Way and external galaxies, as well as cosmology \citep{Nissen2018}. To achieve efficient and robust parameter inference, various surveys have developed parameter-processing pipelines, such as the APOGEE Stellar Parameter and Chemical Abundances Pipeline (ASPCAP; \citealt{GarciaPerez2016}), the data-driven approach to stellar-label determination used in GALAH DR1/DR2 (The Cannon; \citealt{Ness2015,Martell2017,Buder2018}), the Spectroscopy Made Easy adopted in GALAH DR3 (SME; \citealt{Valenti1996,Piskunov2017,Buder2021}), the self-consistent ab initio fitting of stellar spectra used in GALAH DR4 (The Payne; \citealt{Ting2019,Buder2025}), and the General Stellar Parametriser-spectroscopy (GSP-Spec) used for Gaia DR3 RVS (\citealt{RecioBlanco2023}). These pipelines implement the mapping from the stellar-parameter space to the spectral-flux space (spectral emulator), either by invoking physical spectral-synthesis models or by constructing data-driven regression models, and they combine such models with optimization methods appropriate to their implementations to iteratively search for the optimal parameter combination that minimizes the discrepancy between the spectrum predicted by the emulator and the observed spectrum.
	
	In practical applications of large-scale spectroscopic parameterization, stellar parameters often exhibit significant correlations, arising from physical couplings in stellar structure and evolution and also reflected in the statistical properties of the observations. Recent studies have further revealed such structural characteristics of the stellar-parameter space in the Galactic disk: elemental abundances display complex correlation patterns \citep{Casey2019,Weinberg2019,Weinberg2022,Ting_2022,Ness2022,Griffith2022,Griffith2025}, and they manifest as rigidly coupled relations in samples of F- and G-type dwarfs, namely that multiple elemental abundances (e.g., [O/H], [Si/H], [Ca/H]) can be expressed linearly by $T_{\mathrm{eff}}$, $\log g$, [Fe/H], and [Mg/H] \citep{Mead2025}. However, existing parameter-inference methods and optimization algorithms typically solve directly in the original parameter space without explicit treatment of such correlations. In the tests made for this study, we observe that when strong correlations exist among parameters, the optimization process more readily exhibits reduced convergence speed and degraded inference accuracy, suggesting that parameter correlations may affect inference performance to some extent.
	
	In the field of high-dimensional optimization, researchers have proposed various strategies to improve efficiency and stability, such as linear embeddings \citep{Wang2016,Raponi2020,Letham2020,Mei2021,Pretsch2024} and nonlinear embeddings \citep{Tripp2020,Antonov2022,Maus2022,Chu2024,Boyar2024}. These methods are mainly applied within optimization frameworks such as Bayesian optimization and evolutionary algorithms. Their core idea is to reduce the effective optimization dimensionality by searching within an embedded space, thereby mitigating, to some extent, high-dimensional optimization challenges arising from parameter correlations or redundant dimensions.
	
	To alleviate the impact of correlations among high-dimensional stellar parameters on inference accuracy and convergence efficiency, and inspired by embedding ideas in high-dimensional optimization, we propose the Projected-Space Inference of Stellar Parameters (PISP): during the inference stage, an orthonormal projection basis is introduced to map stellar parameters to projection coefficients and optimization is then carried out in the coefficient space. The orthonormal basis can be constructed using either principal component analysis (PCA; \citealt{Hotelling1933,Abdi2010}) or the active-subspace (AS) method (\citealt{Constantine2014,Musayeva2024}). The PCA basis is obtained from the covariance matrix of stellar parameters in the training set, aiming to decorrelate parameters in a statistical sense. In contrast, the AS basis is derived from gradients of the spectral emulator evaluated on the training set. It is not constructed to decorrelate parameters but instead identifies the directions in parameter space to which the spectrum is most sensitive and concentrates the optimization on these key directions, thereby partially alleviating the adverse impact of parameter correlations on optimization convergence and inference stability.
	
	The remainder of this paper is organized as follows. Section~\ref{Data introduction} describes the data used in this work. Section~\ref{Methodology} presents the spectral emulators and the PISP methodology. Section~\ref{Data Experiment} determines the hyperparameters of PISP and evaluates its inference performance. Finally, Section~\ref{Conclusions} concludes the work.
	
	\section{Data} \label{Data introduction}
	PISP performs stellar-parameter inference using spectral emulators trained on Kurucz synthetic spectra. To assess its performance, we apply PISP to both Kurucz synthetic spectra and APOGEE DR$17$ observed spectra.
	
	\begin{table}[htbp]
		\centering
		\caption{Parameter Ranges of the Kurucz Synthetic Spectral Grid}
		\begin{tabular*}{0.48\textwidth}{@{\extracolsep{\fill}}lcc}
			\toprule
			Parameter & Minimum & Maximum \\
			\midrule
			$T_\mathrm{eff}$ (K)                                & 3007 & 7997 \\
			$\log g$ (dex)                                      &    0.0 &    5.3 \\
			$v_\mathrm{turb}$ ($\text{km s}^{-1}$)              &    0.0 &    3.0 \\
			$v_\mathrm{macro}$ ($\text{km s}^{-1}$)             &    0.1 &   30.0 \\
			$\mathrm{[C/H]}$ (dex)               &   -2.5 &    0.9 \\
			$\mathrm{[N/H]}$ (dex)               &   -2.5 &    1.7 \\
			$\mathrm{[O/H]}$ (dex)               &   -2.5 &    1.6 \\
			$\mathrm{[Na/H]}$ (dex)              &   -2.5 &    1.7 \\
			$\mathrm{[Mg/H]}$ (dex)              &   -2.5 &    1.7 \\
			$\mathrm{[Al/H]}$ (dex)              &   -2.5 &    1.7 \\
			$\mathrm{[Si/H]}$ (dex)              &   -2.6 &    0.8 \\
			$\mathrm{[P/H]}$ (dex)               &   -2.5 &    1.7 \\
			$\mathrm{[S/H]}$ (dex)               &   -2.5 &    1.7 \\
			$\mathrm{[K/H]}$ (dex)               &   -2.5 &    1.4 \\
			$\mathrm{[Ca/H]}$ (dex)              &   -2.5 &    1.8 \\
			$\mathrm{[Ti/H]}$ (dex)              &   -2.5 &    1.8 \\
			$\mathrm{[V/H]}$ (dex)               &   -2.5 &    1.7 \\
			$\mathrm{[Cr/H]}$ (dex)              &   -2.5 &    1.7 \\
			$\mathrm{[Mn/H]}$ (dex)              &   -2.5 &    1.7 \\
			$\mathrm{[Fe/H]}$ (dex)              &   -1.5 &    0.5 \\
			$\mathrm{[Co/H]}$ (dex)              &   -2.5 &    1.7 \\
			$\mathrm{[Ni/H]}$ (dex)              &   -2.5 &    1.8 \\
			$\mathrm{[Cu/H]}$ (dex)              &   -2.5 &    1.7 \\
			$\mathrm{[Ge/H]}$ (dex)              &   -2.4 &    1.2 \\
			$^{12}\mathrm{C}/^{13}\mathrm{C}$    &    0.0 &  100.0 \\
			\bottomrule
		\end{tabular*}
		\label{KuruczParamRanges}
	\end{table}

	\subsection{Synthetic Spectra from Kurucz Models}
	The synthetic spectra used in this study were provided by Y.-S.~Ting. This dataset was generated following the same procedure as the refined synthetic model grid described by \citet{Ting2019}, but at a larger scale, comprising $14{,}623$ synthetic spectra that cover a $25$-dimensional stellar-parameter space (\reftable{KuruczParamRanges}); each spectrum consists of $7214$ flux pixels. We divided the dataset into a training set ($12{,}000$ spectra), a validation set ($1312$ spectra), and a test set ($1311$ spectra), for training the spectral emulator and evaluating both its flux-reconstruction error and the performance of PISP on synthetic spectra.
	
	We emphasize that this dataset is not a regular Cartesian grid in the 25-dimensional stellar-parameter space but was instead constructed using the two-stage sampling strategy described in \citet[Section~2.3]{Ting2019}. This strategy concentrates the synthetic spectral calculations in the local neighborhood of stellar-parameter space that is actually populated by real stars, thereby avoiding the exponential growth in model numbers inherent to a uniformly sampled high-dimensional grid and enabling stable spectral interpolation and parameter inference within this effective parameter domain.
	
	The models were computed with ATLAS$12$ and SYNTHE under a one-dimensional local thermodynamic equilibrium framework. The atmospheric structure is divided into $80$ Rosseland optical-depth layers, with a maximum Rosseland depth of $\tau_{R,\mathrm{max}} {=} 1000$. Numerical convergence in each atmospheric layer is automatically verified. The reference abundances adopt the solar composition of \citet{Asplund2009} and the stellar parameters of Arcturus from \citet{Ramirez2011}, which employs an improved H-band atomic line list whose strong-line parameters are calibrated against solar and Arcturus observations. After convergence of the model atmospheres, synthetic spectra are first computed at a nominal resolution of $R {=} 300{,}000$, then degraded to APOGEE’s observational resolution by convolution with APOGEE’s average line-spread function, and finally normalized following \citet{Ness2015}.
	
	\subsection{APOGEE Survey Data} \label{SurveyData}
	APOGEE is one of the programs in SDSS-III \citep{Majewski2017} and SDSS-IV \citep{Blanton2017}. It employs a near-infrared spectrograph with a  resolving power of $R{\sim}22{,}500$ over a wavelength range of $1.51$--$1.70~\mu\mathrm{m}$. The primary targets are red giants in the Milky Way, with additional red giants in the Large Magellanic Cloud and other nearby dwarf galaxies, as well as a large number of FGKM dwarfs \citep{Smith2021}.
	
	In the seventeenth data release (DR$17$), APOGEE includes $657{,}135$ unique targets and approximately $2.6$ million individual visit spectra, and provides visit-combined spectra produced using pixel-based weighting and a more global weighting\footnote{\url{https://www.sdss4.org/dr17/irspec/spectral\_combination}}. DR17 reports atmospheric parameters, radial velocities, and chemical abundances for more than $20$ elements \citep{Abdurrouf2022}.
	
	In this work, using the official parameter catalog as the reference, we infer atmospheric parameters and elemental abundances for $722{,}896$\footnote{The official catalog contains $733{,}901$ entries; we exclude $10{,}996$ empty FITS entries and nine spectra that failed normalization, yielding $722{,}896$ visit-combined spectra.} visit-combined spectra. The inferred parameters are compared with the official catalog to evaluate the reliability of PISP on observed data.
	
	\subsection{Data Preprocessing} \label{Data Preprocessing}
	Before using PISP to infer high-dimensional parameters from APOGEE stellar spectra, we perform the following preprocessing steps:

	\begin{enumerate} 
		\item \textbf{Spectral normalization.}
		To reduce shape differences between APOGEE and synthetic spectra, we normalize the APOGEE spectra prior to inference following \cite{Ness2015}. This method, based on the data-driven model The Cannon, selects wavelength pixels with minimal response to stellar labels, fits a fourth-order polynomial separately in each of APOGEE’s three spectral bands to approximate the pseudocontinuum, and finally divides the flux by this pseudocontinuum to obtain the normalized spectrum.
		\item \textbf{Masking outlier pixels.}
		To ensure reliable spectral fitting, we adopt a dual-mask strategy to exclude abnormal pixels. First, we directly adopt the mask table from \cite{Ting2019}, constructed by comparing synthetic spectra of the Sun and Arcturus with APOGEE observations, which flags pixels with differences between the normalized synthetic and observed spectra exceeding $2\%$. In parallel, we apply the official APOGEE mask to remove instrument- and observation-related bad pixels. Taking the union of the two yields a per-pixel binary mask $A_{\lambda_i}{\in}\{0,1\}$, where $A_{\lambda_i}{=}0$ denotes a bad pixel, $A_{\lambda_i}{=}1$ denotes a valid pixel, and $\lambda_i$ denotes the $i$th wavelength pixel.
	\end{enumerate}

	\section{Methodology} \label{Methodology}
	
	This section first introduces the spectral emulators used for parameter inference (\refsubsection{Spectral-Emulator}), then details the steps of PISP (\refsubsection{PISP-method}), and finally presents a baseline strategy---direct optimization in the standardized stellar-parameter space---to evaluate the effectiveness of PISP (\refsubsection{RawOptimization}).
	
	\subsection{Spectral Emulator} \label{Spectral-Emulator}
	To assess the behavior of PISP with different types of spectral emulators, we construct two architectures: a fully connected neural network (FNN)-based emulator and a residual network (ResNet)-based emulator.
	\begin{itemize}
		\item \textbf{FNN emulator.} The input is a $25$-dimensional stellar-label vector and the output is a $7214$-dimensional APOGEE flux vector. The network has three hidden layers (with $256$, $512$, and $1024$ units, respectively); all hidden layers use ReLU activations and the output layer is linear.
		\item \textbf{ResNet emulator.} The input is a $25$-dimensional stellar-label vector and the output is a $7214$-dimensional APOGEE flux vector. The architecture comprises three fully connected layers and seven deconvolution layers, with residual connections in layers two to six. This implementation follows an open-source reference\footnote{\url{https://github.com/tingyuansen/The_Payne/blob/master/The_Payne/training.py}} and is used to verify the applicability of PISP across network architectures.
	\end{itemize}
	
	Both emulators share the same training strategy. First, each of the $25$ input parameters is scaled independently to the interval $[-0.5, 0.5]$ to alleviate numerical instabilities caused by differences in parameter scales \citep{Nocedal2006}. Specifically,
	\begin{equation}
		\theta {=} \frac{\theta_{\text{raw}} {-} \min(\theta_{\text{raw}}^{\text{train}})}{\max(\theta_{\text{raw}}^{\text{train}}) {-} \min(\theta_{\text{raw}}^{\text{train}})} {-} 0.5,
		\label{norm}
	\end{equation}
	where $\theta_{\text{raw}}$ denotes the raw stellar parameters, $\max(\theta_{\text{raw}}^{\text{train}})$ and $\min(\theta_{\text{raw}}^{\text{train}})$ are the per-dimension maxima and minima computed on the training set, respectively, and $\theta$ is the normalized parameter vector. We use the mean absolute error (MAE) as the loss function; all models are trained for up to $10^{5}$ epochs with Adam for parameter updates. In addition, we employ a plateau learning-rate scheduler (ReduceLROnPlateau)\footnote{\url{https://pytorch.org/docs/stable/generated/torch.optim.lr_scheduler.ReduceLROnPlateau.html}} based on the validation-set MAE. If the validation MAE does not improve for $500$ consecutive epochs, the learning rate is reduced by a factor of $0.9$, with a cooldown of $200$ epochs, down to a minimum of $10^{-8}$. If, after reaching the minimum learning rate, the validation MAE still fails to improve for $5000$ consecutive epochs, training is stopped early to avoid overfitting. Finally, we save the weights corresponding to the lowest validation MAE as the final configuration for each emulator, which are then used for subsequent PISP inference optimization. This unified training framework ensures comparability between the two emulators and provides a basis for evaluating the effectiveness of PISP.
	
	\subsection{PISP} \label{PISP-method}
	Given the widespread use of CurveFit (\texttt{scipy.optimize.curve\_fit}; \citealt{Virtanen2020}) and Adam \citep{Kingma2014} in stellar-parameter inference \citep{Ness2015,Ting2019,Xiang2019,Imig2022,Zhang2023,Rozanski2025,Liang2026}, we implement two algorithmic variants to accommodate different computational scenarios: PISP-CurveFit and PISP-Adam. PISP-CurveFit targets CPU environments and is suitable for fast inversion of individual spectra or small-scale tasks; PISP-Adam is mainly implemented based on PyTorch \citep{Paszke2019}, targets GPU environments, and supports large-scale parallelism and batch processing to improve throughput and convergence efficiency. Each variant includes four inference strategies: PCA-Non-L1 and AS-Non-L1 optimize the projection coefficients under a user-specified projected dimensionality; while PCA-L1 and AS-L1 introduce L1 regularization in the full projected space to enable adaptive selection of projection directions. PISP is invoked only at the inference stage and can act directly on pretrained spectral emulators without additional training; while improving inference accuracy, it effectively reduces computational cost.
	
	\subsubsection{PISP-CurveFit} \label{PISP-CurveFit}
	PISP-CurveFit is mainly used for fast inference of stellar parameters from single spectra and can be combined with the Joblib library \citep{Varoquaux2024} for large survey datasets. The method consists of the following steps:
	\begin{enumerate}
		\item \textbf{Single-spectrum storage.}
		To support multicore parallel processing, each normalized observed spectrum from \refsubsection{Data Preprocessing} and its corresponding official pixel mask are stored together in a single \texttt{.npz} file. This format allows fast loading via the NumPy library \citep{Harris2020}, improving I/O throughput and parallel performance.
		
		\item \textbf{Reconstruct standardized stellar parameters.}
		PISP provides two orthonormal-basis constructions for reconstructing the standardized stellar-parameter vector, based on PCA and the AS method. The optimal construction can be selected for a given target dataset (the basis-selection strategy is described in \refsubsection{n-alpha}). To present both within a unified framework, we write the reconstruction in the standardized parameter space as
		 \begin{equation}
		 	\theta_{i} {=} w_{i,1:n} {\cdot} z_{1:n}^{\mathrm{T}} {+} \mu_{\theta}.
		 	\label{PISP-CurveFit-PCA-2}
		 \end{equation}
		Here, $\theta_{i}$ is the standardized stellar-parameter vector ($1 \times 25$) of the $i$th spectrum in an arbitrary dataset; $z$ is the basis matrix ($25 \times 25$); $z_{1:n}$ denotes the first $n$ columns of $z$ ($25 \times n$), and thus $z_{1:n}^{\mathrm{T}}$ has dimension $n \times 25$; $w_{i,1:n}$ is the coefficient vector to be optimized ($1 \times n$); and $\mu_\theta$ is the reference center vector ($1 \times 25$). For PCA, $\mu_\theta$ is the mean vector of the standardized training-set parameters (denoted by $\theta^{\text{tr}}$) along each dimension; for AS, $\mu_{\theta}{=}0$.
		We treat $n$ as a hyperparameter that controls the projected dimensionality: $n{<}25$ corresponds to reconstruction in a lower-dimensional subspace, whereas $n{=}25$ corresponds to an orthogonal reparameterization without dimensionality reduction. The choice and discussion of $n$ are provided in \refsubsection{n-alpha}.
		The basis matrix $z$ is precomputed once on the training set and is then fixed. For PCA, $z$ is formed by the eigenvectors of the sample covariance matrix of $\theta^{\text{tr}}{-}\mu_{\theta}$, ordered by descending eigenvalues; we implement this procedure using \texttt{sklearn.decomposition.PCA} \citep{Pedregosa2011}.
		For AS, we treat the spectral emulator $g(\cdot)$ as a vector-valued function and adopt the shared active subspace for vector-valued functions proposed by \citet[Section~3.1.2]{Musayeva2024} to construct $z$. Specifically, for the standardized parameters of the $k$th training sample, $\theta_{k}^{\text{tr}}$, we define the pixel-averaged sensitivity vector
		\begin{equation*}
			s_k = \frac{1}{m_1}\sum_{j=1}^{m_1}\left.\frac{\partial g(\theta)_j}{\partial \theta}\right|_{\theta=\theta_{k}^{\text{tr}}},
		\end{equation*}
		where $m_1$ is the flux dimensionality ($m_1{=}7214$), $j$ is the wavelength index, and $s_k$ is a row vector of dimension $1 \times 25$. The partial derivatives are computed via \texttt{torch.func.jacrev}. The pixel-averaged sensitivity vector reflects the global spectral sensitivity across all wavelength pixels, where contributions from near-zero gradient pixels and cancellation between pixels with gradients of opposite signs are naturally incorporated, rather than localized sensitivity at individual lines. This averaging step trades off local detail for global robustness, which may reduce the influence of parameters that affect only a few pixels. We further construct
		\begin{equation*}
			C = \frac{1}{m_2}\sum_{k=1}^{m_2} s_k^{\mathrm{T}} s_k,
		\end{equation*}
		 where $m_2$ is the number of training samples ($m_2{=}12{,}000$). This summation provides an empirical estimate of the expected outer product of the pixel-averaged sensitivity vector with itself over the training set. Finally, we perform an eigen-decomposition of $C$ using \texttt{torch.linalg.eigh}, order the eigenvectors by descending eigenvalues, and use them to form the basis matrix $z$.
		 
		\item \textbf{Define the objective function.}
		For the $i$th spectrum $F_i$ ($1{\times}7214$) and its binary pixel-mask vector $A_i$ ($1{\times}7214$), we define the objective function in the standardized parameter space as
		\begin{equation}
			\mathcal{L}_{\scriptscriptstyle \text{CurveFit}}^{\scriptscriptstyle \text{raw}}
			 {=} \sum_{j} \left( F_{i,j} {\cdot} A_{i,j} {-} g(\theta_{i})_{j}  {\cdot} A_{i,j} \right)^2,
			\label{PISP-CurveFit-PCA-loss1}
		\end{equation}
		where $j$ indexes wavelength pixels and $g(\cdot)$ denotes the spectral emulator. Substituting \refformula{PISP-CurveFit-PCA-2} into \refformula{PISP-CurveFit-PCA-loss1} and adding an L1 regularization term yields the objective function with respect to the basis coefficients
		\begin{equation}
			\begin{split}
				\mathcal{L}_{\scriptscriptstyle \text{CurveFit}}^{\scriptscriptstyle \text{PISP}}
				& {=} \sum_{j} (F_{i,j} {\cdot} A_{i,j} {-} g(w_{i,1:n} {\cdot} z_{1:n}^{\mathrm{T}} {+} \mu_{\theta})_{j}  {\cdot} A_{i,j})^2 \\
				&{+} \alpha {\cdot} \sum_{t}|w_{i,t}|,
			\end{split}
			\label{PISP-CurveFit-PCA-loss2}
		\end{equation}
		where $\alpha{\ge} 0$ is the regularization coefficient, and $w_{i,t}$ is the $t$th basis coefficient ($t{=}1,{\cdots},n$). If $\alpha{=}0$, the method corresponds to the PCA-Non-L1 and AS-Non-L1 strategies, in which only the first $n$ coefficients $w_{i,1:n}$ are optimized ($n$ is a hyperparameter, determined in \refsubsection{n-alpha}). If $\alpha{>}0$, the method corresponds to the PCA-L1 and AS-L1 strategies, for which we set $n{=}25$ and optimize all coefficients $w_{i,1:25}$ ($\alpha$ is the hyperparameter of this strategy, also determined in \refsubsection{n-alpha}). The PCA-Non-L1 and AS-Non-L1 strategies perform truncation, discarding the $25{-}n$ low-variance principal components or the $25{-}n$ inactive directions, whereas the PCA-L1 and AS-L1 strategies induce sparsity across the full set of 25 basis directions via regularization. The purpose of designing these four strategies is to evaluate how the basis choice and constraint affect inference accuracy, and to adapt to different datasets.
		
		\item \textbf{Minimization of the objective function.}
		For the $i$th spectrum $F_i$, PISP-CurveFit provides four inference strategies: PCA-Non-L1 ($\alpha{=}0$), PCA-L1 ($\alpha{>}0$), AS-Non-L1 ($\alpha{=}0$), and AS-L1 ($\alpha{>}0$). All four strategies iteratively solve for the minimizer of the objective function in \refformula{PISP-CurveFit-PCA-loss2} using the trust region reflective (trf) algorithm \citep{Coleman1996,Branch1999} via CurveFit. The initial values are uniformly set to the basis coefficients computed via \refformula{PISP-CurveFit-PCA-2} from the standardized stellar parameters corresponding to the nearest-neighbor spectrum of $F_i$ in the training set\footnote{The nearest-neighbor search is implemented using \path{sklearn.neighbors.KNeighborsRegressor(n_neighbors=1)}.}. Following \citet{Ting2019}, we set the convergence tolerances (ftol and xtol) to $5{\times}10^{-4}$. Upon convergence, the optimal basis coefficients $w_{i,1:n}^\ast$ are substituted into \refformula{PISP-CurveFit-PCA-2} to obtain the standardized stellar parameters, which are then transformed back to physical units via the inverse transformation of \refformula{norm} to yield the final stellar parameters.
		
		\item \textbf{Estimating parameter errors.}
		We define the residual row vector function for the $i$th spectrum $F_i$ as
		\begin{equation}
			{r}(w_{i,1:n}) {=}
			\begin{cases}
				\Delta f_i, & (\text{\small a})\\
				\begin{bmatrix}
					(\Delta f_i)^{\mathrm{T}} \\
					-\sqrt{\alpha|w_{i,1}|} \\
					{\vdots} \\ 
					-\sqrt{\alpha|w_{i,n}|}
				\end{bmatrix}^{\mathrm{T}}, & (\text{\small b})
			\end{cases}
			\label{curvefit_residual}
		\end{equation}
		where (a) denotes PCA-Non-L1 and AS-Non-L1, and (b) denotes PCA-L1 and AS-L1. Here, $\Delta f_i$ is the row vector formed by extracting elements with mask $A_{i,j}{=}1$ from $F_i - g(w_{i,1:n} \cdot z_{1:n}^{\mathrm{T}} + \mu_\theta)$ in wavelength order ($1 \times d$, where $d=\sum_j A_{i,j}$). Once the optimal basis coefficients $w_{i,1:n}^\ast$ of \refformula{PISP-CurveFit-PCA-loss2} are obtained, we compute the Jacobian matrix of \refformula{curvefit_residual} at $w_{i,1:n}^\ast$ using the two-point finite difference approximation in the \texttt{SciPy} library
		\begin{equation}
			J_i {=} \left.\frac{\partial r(w_{i,1:n})}{\partial w_{i,1:n}}\right|_{w_{i,1:n}=w_{i,1:n}^\ast}.
			\label{J_i}
		\end{equation}
		We then estimate the parameter covariance matrix using the linear approximation \citep{Vugrin2007}
		\begin{equation}
			\Sigma_i {=}
			\begin{cases}
				\displaystyle
				\frac{r(w_{i,1:n}^\ast)\,r(w_{i,1:n}^\ast)^{\mathrm{T}}}{d-n}\left(J_i^{\mathrm{T}}J_i\right)^{-1}, & (\text{\small a})\\[6pt]
				\displaystyle
				\frac{r(w_{i,1:n}^\ast)\,r(w_{i,1:n}^\ast)^{\mathrm{T}}}{d}\left(J_i^{\mathrm{T}}J_i\right)^{-1}, & (\text{\small b})
			\end{cases}
			\label{cov_x}
		\end{equation}
		and according to \refformula{norm} and \refformula{PISP-CurveFit-PCA-2}, we transform \refformula{cov_x} into the covariance matrix of the original stellar parameters
		\begin{equation*}
			\Sigma_i' {=}
			\begin{cases}
				Qz_{1:n}\Sigma_iz_{1:n}^{\mathrm{T}}Q^{\mathrm{T}}, & (\text{\small a})\\
				Qz\Sigma_iz^{\mathrm{T}}Q^{\mathrm{T}}, & (\text{\small b})
			\end{cases}
		\end{equation*}
		where $Q$ is the diagonal matrix formed by $\max(\theta_{\text{raw}}^{\text{train}})-\min(\theta_{\text{raw}}^{\text{train}})$ in \refformula{norm}. Finally, the stellar-parameter errors for the $i$th spectrum are the square roots of the diagonal elements of the covariance matrix $\Sigma_i'$.
	\end{enumerate}
	
	For large survey applications, we control the number of concurrent processes in Joblib via the parameter \texttt{n\_jobs}. Following Figure~1 of \cite{Liang2026}, we set \texttt{n\_jobs=-1}, i.e., using all available CPU cores; this setting exhibits near-optimal inference efficiency across various hardware configurations.
	
	\subsubsection{PISP-Adam}\label{PISP-Adam}
	PISP-Adam is designed for stellar-parameter inference in large spectroscopic datasets. Its overall workflow comprises the following steps:
	\begin{enumerate}
		\item \textbf{Batch spectrum storage.} 
		To accommodate large-scale inference in a GPU environment, every $20{,}000$ spectra from \refsubsection{Data Preprocessing} are packaged into a single PyTorch tensor file (\texttt{.pt}). Each \texttt{.pt} file contains the normalized spectra, the official pixel masks, and the parameter initial values generated in the same way as for PISP-CurveFit. This design enables fast loading of the required data and initial values when jointly inferring $N {\leq} 20{,}000$ spectra with $25$ stellar parameters each, facilitating subsequent optimization. During inference, to balance computational efficiency and GPU memory usage, the default $N$ is set according to the type of spectral emulator: $N{=}10{,}000$ for the FNN emulator and $N{=}5000$ for the ResNet emulator. This setting follows the relationship between $N$ and inference efficiency shown in Figure~1 of \cite{Liang2026}; throughput is typically higher when $N$ approaches the upper bound of available GPU memory capacity.
		
		\item \textbf{Batch reconstruction of standardized stellar parameters.}
    	From \refformula{PISP-CurveFit-PCA-2}, we obtain the batch reconstruction of standardized stellar parameters for any $N$ spectra
		\begin{equation}
			\theta_{\text{Adam}} {=}
			\begin{pmatrix}
				\theta_{1} \\
				\vdots\\
				\theta_{N}
			\end{pmatrix}
			{=}
			\begin{pmatrix}
				w_{1,1:n} {\cdot} z_{1:n}^{\mathrm{T}} {+} \mu_{\theta}\\
				\vdots\\
				w_{N,1:n} {\cdot} z_{1:n}^{\mathrm{T}} {+} \mu_{\theta}\\
			\end{pmatrix},
			\label{PISP-Adam-PCA}
		\end{equation}
    	where $\theta_{\text{Adam}}$ is the reconstructed $N \times 25$ matrix of standardized stellar parameters. The reconstruction is implemented in PyTorch with batched computation to improve efficiency.
    	
    	\begin{figure*}[htp!]
    		\centering
    		\includegraphics[width=\textwidth,height=0.7\textheight, keepaspectratio]{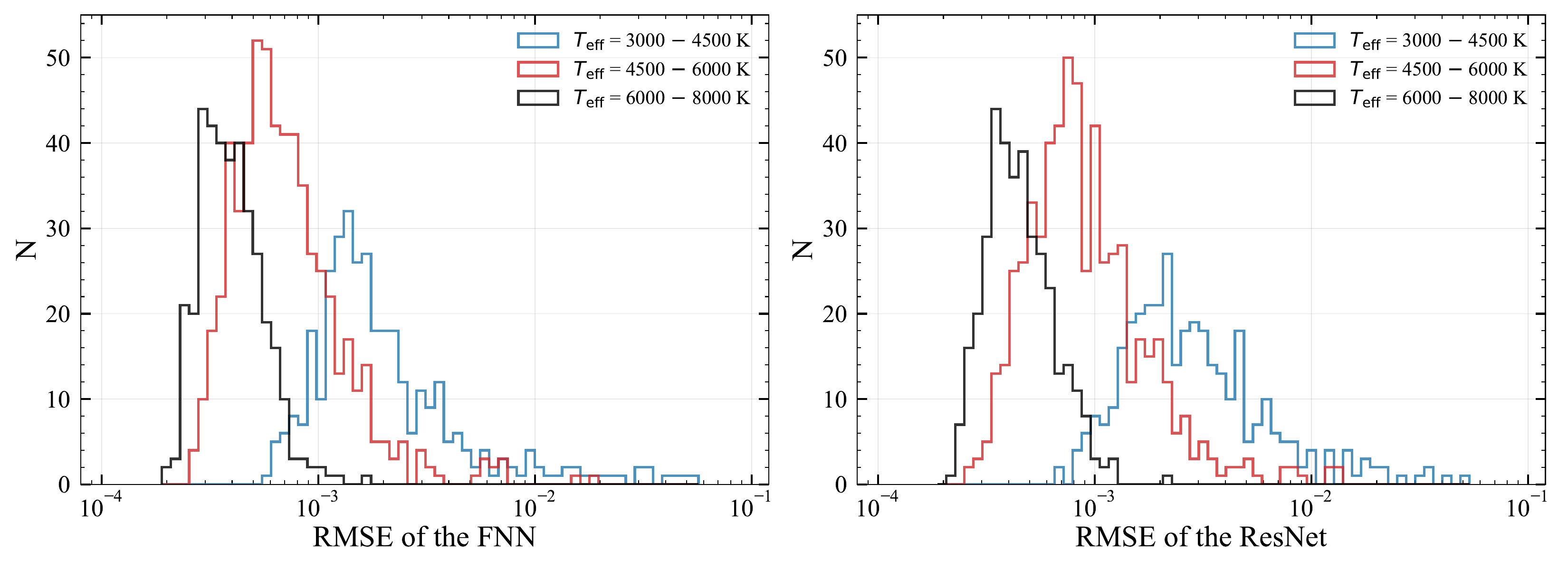}
    		\caption{Distribution of spectral flux-reconstruction errors (RMSE) across different $T_{\rm eff}$ ranges. The left and right panels show the RMSE histograms of the FNN- and ResNet-based spectral emulators on the test set, respectively.}
    		\label{ANN_ResNet_flux_reproduced}
    	\end{figure*}
    	
		\item \textbf{Define the objective function.}
		For any $N$ spectra $F{=}(F_{1}, {\dots}, F_{N})$ ($N {\times} m_1$, where $m_1{=}7214$ is the total flux dimensionality) and their pixel-mask matrix $A{=}(A_{1}, {\dots}, A_{N})$ ($N {\times} m_1$), we define the objective function in the standardized stellar-parameter space
		{\small
			\begin{equation}
				\mathcal{L}_{\scriptscriptstyle \text{Adam}}^{\scriptscriptstyle \text{raw}}
				{=} \sqrt{\frac{1}{N {\cdot} m_1} \sum_{i,j} \left( F_{i,j} {\cdot} A_{i,j} {-} g(\theta_{\text{Adam}})_{i,j} {\cdot} A_{i,j} \right)^2},
				\label{PISP-Adam-PCA-loss1}
		\end{equation}}
	
		where
		\begin{equation*}
			g(\theta_{\text{Adam}}) {=}
			\begin{pmatrix}
				g(\theta_{1}) \\
				\vdots\\
				g(\theta_{N})
			\end{pmatrix},
		\end{equation*}
		and $i,j$ index the $j$th wavelength pixel of the $i$th spectrum. Substituting \refformula{PISP-Adam-PCA} into \refformula{PISP-Adam-PCA-loss1} and adding L1 regularization yields the objective function in terms of basis coefficients:
		{\small
			\begin{equation}
				\begin{split}
					\mathcal{L}_{\scriptscriptstyle \text{Adam}}^{\scriptscriptstyle \text{PISP}}
					&{=} \sqrt{\frac{1}{N {\cdot} m_1} \sum_{i,j} (e_{i,j})^2 {+} \frac{\alpha}{N {\cdot} m_1} {\cdot} \sum_{i,t}|w_{i,t}|},
				\end{split}
				\label{PISP-Adam-PCA-loss2}
		\end{equation}}
		where
		\begin{equation*}
			e_{i,j} {=}
			F_{i,j} {\cdot} A_{i,j} {-} g(w_{i,1:n} {\cdot} z_{1:n}^{{\mathrm{T}}} {+} \mu_{\theta})_{j} {\cdot} A_{i,j},
		\end{equation*}
		and $\alpha{\ge} 0$ is the regularization coefficient. As in PISP-CurveFit, when $\alpha{=}0$, the method corresponds to the PCA-Non-L1 and AS-Non-L1 strategies, in which only the first $n$ columns of basis coefficients $w_{i,1:n}$ ($i{=}1,{\cdots},N$) for all $N$ samples are optimized; when $\alpha{>}0$, the method corresponds to the PCA-L1 and AS-L1 strategies, for which we set $n{=}25$ and optimize all basis coefficients $w_{i,1:25}$ ($i{=}1,{\cdots},N$). Here, $n$ is the hyperparameter for the Non-L1 strategies, and $\alpha$ is the hyperparameter for the L1 strategies; their determination is described in \refsubsection{n-alpha}. To improve inference efficiency, both model spectrum generation and objective function computation are implemented using PyTorch for batch parallel processing.
		
		\item \textbf{Minimization of the objective function.}
		PISP-Adam supports four inference strategies: PCA-Non-L1 ($\alpha{=}0$), PCA-L1 ($\alpha{>}0$), AS-Non-L1 ($\alpha{=}0$), and AS-L1 ($\alpha{>}0$). All four strategies iteratively solve for the minimizer of the objective function in \refformula{PISP-Adam-PCA-loss2} using the Adam optimizer \citep{Kingma2014}, implemented via \texttt{torch.optim.Adam}. The default learning rate is set to $0.01$, and the initial values are the same as in PISP-CurveFit. We define three convergence criteria: reaching $10{,}000$ iterations, a change in the objective function smaller than $10^{-8}$ over $50$ consecutive iterations, or a monotonically increasing objective over $50$ consecutive iterations. The learning rate and convergence threshold are hyperparameters of PISP-Adam and are chosen to balance convergence stability and computational efficiency (see \refsection{lr-eps}). Satisfaction of any criterion is taken as convergence, after which the optimal basis coefficients $w_{i,1:n}^{\ast}$ ($i{=}1,{\cdots},N$) are substituted into \refformula{PISP-Adam-PCA} to obtain $\theta_{\text{Adam}}^{\ast}$, which are then transformed back to physical units via the inverse transformation of \refformula{norm}. The computation of parameter errors follows the same procedure as in PISP-CurveFit.
	\end{enumerate}
	
	\subsection{Baseline Strategy: Direct Optimization in the Standardized Stellar-Parameter Space} \label{RawOptimization}
	To evaluate the performance of PISP, we design a baseline strategy: direct optimization in the standardized stellar-parameter space without orthogonal transformation. This strategy serves as the control group for both PISP-CurveFit and PISP-Adam implementations. The optimization variable is uniformly set to the standardized stellar-parameter vector $\theta_i$ (rather than basis coefficients), the objective functions are \refformula{PISP-CurveFit-PCA-loss1} and \refformula{PISP-Adam-PCA-loss1}, respectively, and the optimization algorithms are CurveFit and Adam, respectively.
	
	To ensure a fair comparison, this strategy adopts the same initialization as PISP: the initial values are uniformly set to the standardized stellar parameters corresponding to the nearest-neighbor spectrum of the target spectrum in the training set. The convergence criteria during optimization are also consistent with PISP. The final standardized parameters $\theta_i^{\ast}$ are transformed back to stellar parameters with physical units via the inverse transformation of \refformula{norm}. Parameter errors are computed using the same method as PISP: the Jacobian matrix is computed at the optimum $\theta_i^{\ast}$ using two-point finite differences, and the covariance matrix of the standardized stellar parameters is estimated using linear approximation, which is then transformed to the original parameter space according to \refformula{norm}; the square roots of the diagonal elements are taken as the final stellar-parameter errors.
	
	This direct optimization strategy is not part of the PISP framework itself but is rather a baseline strategy without orthogonal transformation or sparsity constraints, which provides a reference for evaluating PISP's convergence speed and inference accuracy in high-dimensional space. In the following sections, we present experimental comparisons to quantitatively demonstrate the differences between this strategy and PISP in inference accuracy and efficiency.
	
	\begin{figure*}[htp!]
		\centering
		\includegraphics[width=\textwidth,height=0.7\textheight, keepaspectratio]{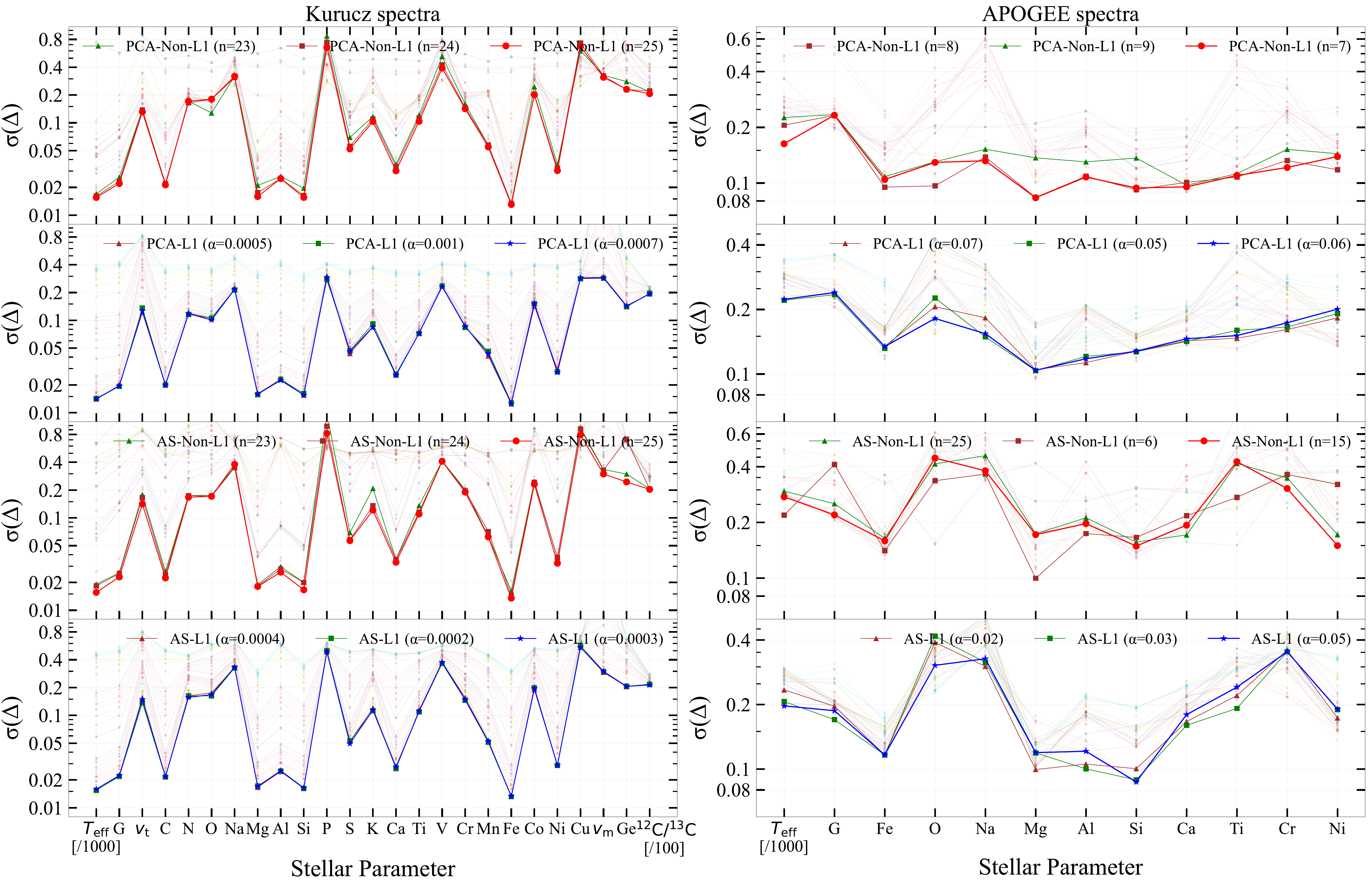}
		\caption{Hyperparameter settings ($n$ and $\alpha$) of PISP across different datasets. The first column lists the hyperparameters tuned on the Kurucz validation set, while the second column shows those optimized using the APOGEE DR$17$ reference set. The x-axis indicates the stellar parameters and the y-axis shows the standard deviation ($\sigma$) of the differences ($\Delta$) between inferred and reference values, with $\sigma$ computed using \texttt{astropy.stats.sigma\_clipped\_stats} (\texttt{sigma=3}, \texttt{maxiters=3}). From top to bottom, the four rows correspond to the PCA-Non-L1, PCA-L1, AS-Non-L1, and AS-L1 strategies, respectively. In the PCA-Non-L1 and AS-Non-L1 panels, the red curve denotes the optimal projection dimension $n$, while in the PCA-L1 and AS-L1 panels, the blue curve denotes the optimal $\alpha$. The lighter curves in each panel show the results for all other tested hyperparameter values (e.g., in the first-row first-column panel, all cases with $n{<}23$ are shown). The y-axis adopts a base-$10$ logarithmic scale, with minor ticks indicating evenly spaced subdivisions between adjacent major ticks and tick labels showing the corresponding linear values. For clarity, $\log g$, $V_{\mathrm{turb}}$, and $V_{\mathrm{macro}}$ are abbreviated as $G$, $V_\mathrm{t}$, and $V_\mathrm{m}$ along the x-axis.}
		\label{valid_params}
	\end{figure*}
	
	\section{Experiment} \label{Data Experiment}
	This section systematically evaluates the performance of PISP on both theoretical and observed spectra. First, we quantify the flux-reconstruction errors of the two types of spectral emulators (FNN and ResNet) on the test set to verify their fitting capabilities. Second, we separately tune PISP's key hyperparameters on the theoretical validation set and the observational reference set and select the optimal projection basis construction method. After determining the optimal settings, we assess PISP's parameter inference accuracy relative to the baseline strategy (direct optimization in the standardized stellar-parameter space) on both theoretical spectra (test set) and APOGEE DR$17$ observed spectra. We examine the validity of the error model by comparing the consistency between model errors and repeat-observation errors and also evaluate whether PISP's inference accuracy depends on the choice of optimizer by testing multiple optimization algorithms. Finally, we assess PISP's computational efficiency and discuss potential directions for improvement.
	
	\subsection{Evaluating the Performance of the Spectral Emulator} \label{emulator-performance}
	To assess the fitting capability of the spectral emulators, we compute, on the test set, the flux-reconstruction error for each normalized spectrum, i.e., the root mean square error (RMSE) of the flux, and we divide the sample by $T_{\rm eff}$ into three bins: $3000$--$4500$ K, $4500$--$6000$ K, and $6000$--$8000$ K (see \reffig{ANN_ResNet_flux_reproduced}). Overall, RMSE decreases with increasing $T_{\rm eff}$. In the low-temperature bin ($3000$--$4500$ K), the median RMSE of the FNN is $0.0016$, lower than the ResNet’s $0.0025$; the fractions with RMSE $<10^{-3}$ are $13.4\%$ and $4.8\%$, respectively, indicating that the FNN is more stable for cool-star spectra. In the intermediate bin ($4500$--$6000$ K), both models perform well, with the FNN retaining a slight advantage (fraction with RMSE $<10^{-3}$: $77.4\%$ versus \ $63.1\%$). In the high-temperature bin ($6000$--$8000$ K), the median RMSE of both models drops to about $0.0004$, and the fraction with RMSE $<10^{-3}$ exceeds $98\%$ for both, indicating essentially comparable reconstruction errors. Given that the two emulators use the same training set, validation strategy, and ReduceLROnPlateau scheduler, and that the final models are selected at the minimum validation loss, the observed differences are unlikely to originate from the training procedure and more likely reflect architectural effects. It is worth noting that the absolute difference between the median RMSEs is less than $10^{-3}$, and no significant difference is observed in subsequent parameter inference. We employ two structurally different emulators---FNN and ResNet---to span different inductive biases and representational capacities, ensuring that the conclusions about PISP do not depend on a single architecture, rather than to compare the emulators themselves.
	
	\begin{table}[htbp]
		\centering
		\caption{Comparison of PCA and AS Projection Bases for PISP with Optimal Hyperparameter Configurations}
		\label{pisp_hyperparams}
		\renewcommand{\arraystretch}{1.3}
		\begin{tabular}{lcccc}
			\hline\hline
			Strategy & Dataset & $n$ & $\alpha$ & $\sum \sigma(\Delta)$ \\
			\hline
			PCA-Non-L1 & Kurucz & 25 & 0      & \textbf{4.12} \\
			AS-Non-L1  & Kurucz & 25 & 0      & 4.60 \\
			\hline
			PCA-L1     & Kurucz & 25 & 0.0007 & \textbf{2.65} \\
			AS-L1      & Kurucz & 25 & 0.0003 & 3.80 \\
			\hline
			PCA-Non-L1 & APOGEE & 7  & 0      & \textbf{1.51} \\
			AS-Non-L1  & APOGEE & 15 & 0      & 3.07 \\
			\hline
			PCA-L1     & APOGEE & 25 & 0.06   & \textbf{1.95} \\
			AS-L1      & APOGEE & 25 & 0.05   & 2.42 \\
			\hline
		\end{tabular}
		
		\medskip  
		\parbox{0.4\textwidth}{\small \textbf{Note.} The minimum $\sum \sigma(\Delta)$ between PCA and AS within each dataset and regularization setting is highlighted in boldface. The definition of $\sigma(\Delta)$ is given in Figure~\ref{valid_params}.}
	\end{table}
	\begin{figure*}[htp!]
		\centering
		\includegraphics[width=\textwidth,height=0.7\textheight, keepaspectratio]{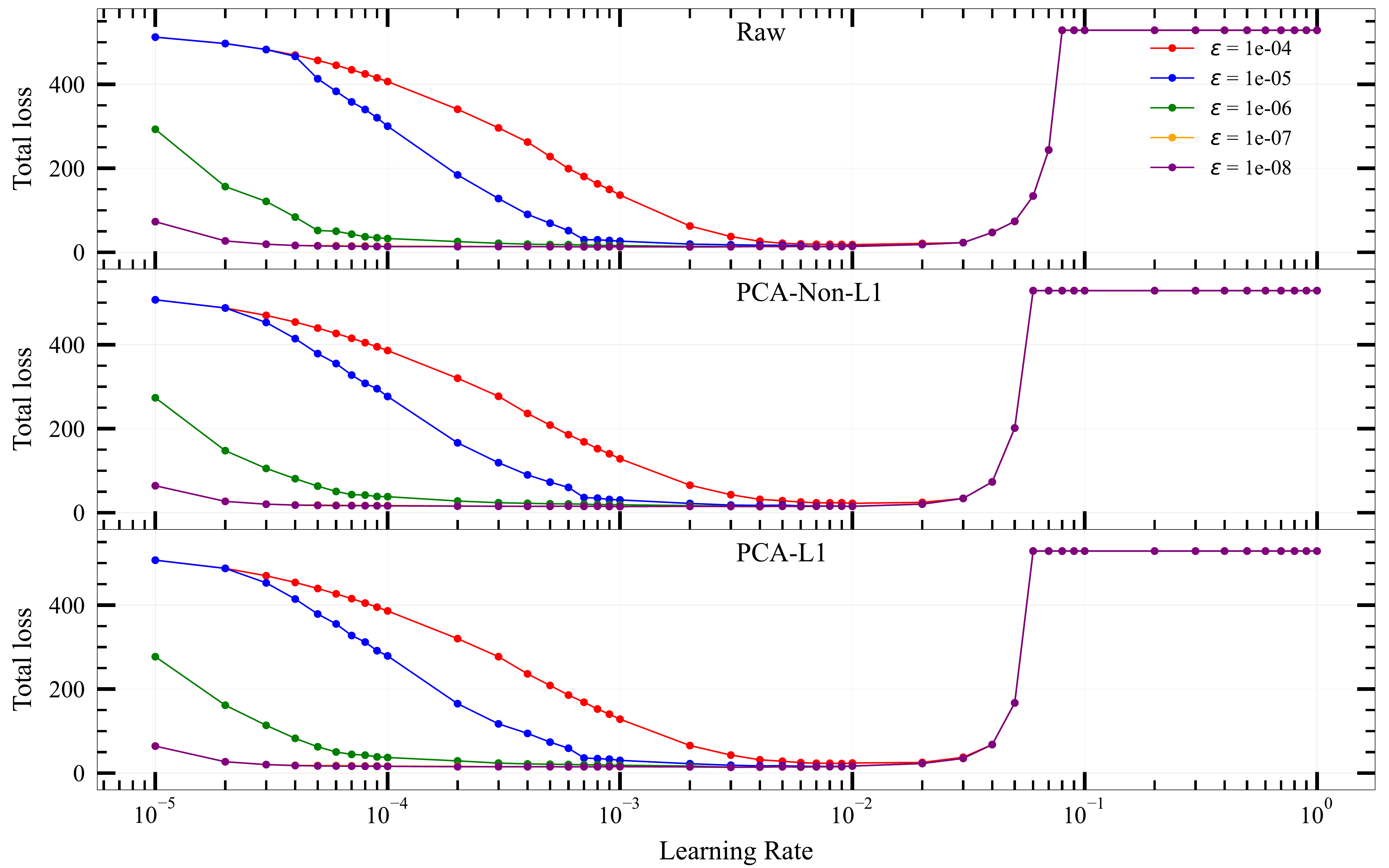}
		\caption{Hyperparameter settings ($lr$ and $\epsilon$) of PISP-Adam on the Kurucz validation set. The three panels correspond to optimization in the baseline strategy (Raw), PCA-Non-L1 ($n{=}25$), and PCA-L1 ($\alpha{=}7 {\times} 10^{-4}$), respectively. Each curve represents a fixed convergence threshold $\epsilon$ ranging from $10^{-8}$ to $10^{-4}$. The x-axis adopts a base-$10$ logarithmic scale to represent the learning rate $lr$, with tick labels indicating the corresponding linear values and minor ticks evenly spaced between adjacent major ticks.}
		\label{Adam_best_lr}
	\end{figure*}
	
	\begin{table*}[htbp]
		\centering
		\caption{Standard Deviations of Stellar Parameter Differences with Respect to Kurucz Reference Labels on the Test Set}
		\label{Kurucz_sigma_emulator_optimizer}
		\setlength{\tabcolsep}{7.5pt}
		\renewcommand{\arraystretch}{1.2}
		\small  
		\begin{tabular}{lcccccccccccc}
			\hline\hline
			\multirow{3}{*}{\centering Parameter} &
			\multicolumn{6}{c}{FNN} &
			\multicolumn{6}{c}{ResNet} \\
			\cmidrule(lr){2-7}\cmidrule(lr){8-13}
			& \multicolumn{3}{c}{CurveFit} & \multicolumn{3}{c}{Adam}
			& \multicolumn{3}{c}{CurveFit} & \multicolumn{3}{c}{Adam} \\
			\cmidrule(lr){2-4}\cmidrule(lr){5-7}\cmidrule(lr){8-10}\cmidrule(lr){11-13}
			& Raw & Non-L1 & L1 & Raw & Non-L1 & L1 & Raw & Non-L1 & L1 & Raw & Non-L1 & L1 \\
			\hline
			$T_{\rm eff}$ & 16 & 16 & \textbf{14} & 16 & 16 & \textbf{15} & 15 & 15 & \textbf{14} & 15 & 15 & \textbf{14} \\
			$\log g$ & 0.03 & \textbf{0.02} & \textbf{0.02} & \textbf{0.02} & \textbf{0.02} & \textbf{0.02} & \textbf{0.02} & \textbf{0.02} & \textbf{0.02} & \textbf{0.02} & \textbf{0.02} & \textbf{0.02} \\
			$v_{\rm turb}$ & 0.14 & \textbf{0.13} & \textbf{0.13} & 0.14 & \textbf{0.13} & \textbf{0.13} & \textbf{0.12} & 0.13 & 0.13 & 0.13 & \textbf{0.12} & 0.13 \\
			{[C/H]} & \textbf{0.02} & \textbf{0.02} & \textbf{0.02} & 0.03 & \textbf{0.02} & \textbf{0.02} & \textbf{0.02} & \textbf{0.02} & \textbf{0.02} & \textbf{0.02} & \textbf{0.02} & \textbf{0.02} \\
			{[N/H]} & 0.19 & 0.17 & \textbf{0.11} & 0.20 & 0.17 & \textbf{0.11} & 0.15 & 0.14 & \textbf{0.10} & 0.15 & 0.13 & \textbf{0.10} \\
			{[O/H]} & 0.22 & 0.16 & \textbf{0.09} & 0.20 & 0.14 & \textbf{0.10} & 0.14 & 0.12 & \textbf{0.08} & 0.13 & 0.12 & \textbf{0.08} \\
			{[Na/H]} & 0.37 & 0.30 & \textbf{0.21} & 0.34 & 0.29 & \textbf{0.22} & 0.28 & 0.22 & \textbf{0.15} & 0.27 & 0.22 & \textbf{0.16} \\
			{[Mg/H]} & 0.02 & 0.02 & \textbf{0.01} & \textbf{0.02} & \textbf{0.02} & \textbf{0.02} & \textbf{0.01} & \textbf{0.01} & \textbf{0.01} & \textbf{0.01} & \textbf{0.01} & \textbf{0.01} \\
			{[Al/H]} & 0.03 & 0.03 & \textbf{0.02} & \textbf{0.03} & \textbf{0.03} & \textbf{0.03} & \textbf{0.02} & \textbf{0.02} & \textbf{0.02} & \textbf{0.02} & \textbf{0.02} & \textbf{0.02} \\
			{[Si/H]} & 0.02 & 0.02 & \textbf{0.01} & 0.02 & 0.02 & \textbf{0.01} & 0.02 & 0.02 & \textbf{0.01} & 0.02 & \textbf{0.01} & \textbf{0.01} \\
			{[P/H]} & 1.02 & 0.72 & \textbf{0.30} & 0.72 & 0.42 & \textbf{0.31} & 0.88 & 0.68 & \textbf{0.32} & 0.82 & 0.47 & \textbf{0.31} \\
			{[S/H]} & 0.06 & \textbf{0.05} & \textbf{0.05} & 0.07 & 0.06 & \textbf{0.05} & 0.05 & 0.05 & \textbf{0.04} & 0.05 & 0.05 & \textbf{0.04} \\
			{[K/H]} & 0.11 & 0.10 & \textbf{0.09} & 0.15 & 0.11 & \textbf{0.09} & 0.07 & \textbf{0.06} & 0.07 & \textbf{0.07} & \textbf{0.07} & \textbf{0.07} \\
			{[Ca/H]} & \textbf{0.03} & \textbf{0.03} & \textbf{0.03} & 0.04 & 0.04 & \textbf{0.03} & 0.03 & 0.03 & \textbf{0.02} & 0.03 & 0.03 & \textbf{0.02} \\
			{[Ti/H]} & 0.11 & 0.10 & \textbf{0.07} & 0.14 & 0.10 & \textbf{0.08} & 0.09 & 0.09 & \textbf{0.07} & 0.10 & 0.09 & \textbf{0.07} \\
			{[V/H]} & 0.49 & 0.34 & \textbf{0.22} & 0.40 & 0.33 & \textbf{0.24} & 0.56 & 0.44 & \textbf{0.24} & 0.52 & 0.35 & \textbf{0.25} \\
			{[Cr/H]} & 0.17 & 0.14 & \textbf{0.08} & 0.14 & 0.12 & \textbf{0.09} & 0.09 & 0.10 & \textbf{0.07} & 0.10 & 0.09 & \textbf{0.07} \\
			{[Mn/H]} & 0.05 & 0.05 & \textbf{0.04} & 0.06 & 0.05 & \textbf{0.04} & \textbf{0.04} & \textbf{0.04} & \textbf{0.04} & 0.05 & 0.05 & \textbf{0.04} \\
			{[Fe/H]} & \textbf{0.01} & \textbf{0.01} & \textbf{0.01} & \textbf{0.01} & \textbf{0.01} & \textbf{0.01} & \textbf{0.01} & \textbf{0.01} & \textbf{0.01} & \textbf{0.01} & \textbf{0.01} & \textbf{0.01} \\
			{[Co/H]} & 0.22 & 0.19 & \textbf{0.14} & 0.23 & 0.19 & \textbf{0.14} & 0.30 & 0.28 & \textbf{0.17} & 0.31 & 0.24 & \textbf{0.17} \\
			{[Ni/H]} & \textbf{0.03} & \textbf{0.03} & \textbf{0.03} & \textbf{0.03} & \textbf{0.03} & \textbf{0.03} & 0.05 & 0.04 & \textbf{0.03} & 0.05 & 0.04 & \textbf{0.03} \\
			{[Cu/H]} & 0.97 & 0.77 & \textbf{0.29} & 0.75 & 0.36 & \textbf{0.29} & 0.39 & 0.32 & \textbf{0.23} & 0.39 & 0.29 & \textbf{0.23} \\
			$v_{\rm macro}$ & \textbf{0.33} & 0.35 & 0.34 & 0.36 & 0.35 & \textbf{0.34} & 0.34 & 0.34 & \textbf{0.31} & 0.36 & 0.34 & \textbf{0.31} \\
			{[Ge/H]} & 0.21 & 0.17 & \textbf{0.14} & 0.20 & 0.19 & \textbf{0.14} & 0.15 & 0.15 & \textbf{0.13} & 0.16 & 0.14 & \textbf{0.13} \\
			$^{12}{\rm C}/^{13}{\rm C}$ & 23 & 21 & \textbf{19} & 20 & 19 & \textbf{18} & 27 & 22 & \textbf{18} & 22 & 19 & \textbf{18} \\
			\hline
		\end{tabular}
		
		\vspace{0.3cm}
		\parbox{\textwidth}{\small \textbf{Note.} Values are $\sigma$ computed using \texttt{astropy.stats.sigma\_clipped\_stats} (\texttt{sigma=3}, \texttt{maxiters=3}) for differences ($\Delta{=}\mathrm{inferred}{-}\mathrm{reference}$). For brevity, `Raw' denotes the baseline strategy, `Non-L1' refers to PCA-Non-L1 ($n{=}25$), and `L1' refers to PCA-L1 ($\alpha{=}0.0007$). Within each emulator--optimizer pair, the minimum $\sigma$ among the three strategies is highlighted in boldface.}
	\end{table*}

	\subsection{Configuration of PISP} \label{pisp-hyperopt}
	PISP requires a predefined set of projection bases for the inference stage and performs optimization within the corresponding projection space. Consequently, the parameter inference results depend on both the basis construction method and the associated hyperparameter settings. This work implements and compares two basis construction methods: PCA-based and AS-based. Beyond the choice of basis construction, PISP's configuration parameters can be functionally categorized into two groups: (1) projection-space-related parameters, including the number of retained projection dimensions $n$ in the PCA-Non-L1 and AS-Non-L1 strategies, and the regularization coefficient $\alpha$ in the PCA-L1 and AS-L1 strategies, which suppress the influence of certain projection directions on parameter inference and require independent tuning for different datasets; and (2) optimizer-related parameters, including the learning rate $lr$ and convergence threshold $\epsilon$, which primarily affect the stability and convergence speed of the iterative process and generally exhibit strong transferability across datasets. Based on this division, we adopt a two-stage tuning strategy. In the first stage, using the Kurucz theoretical spectrum validation set and the APOGEE DR$17$ reference subset (see \refsubsection{n-alpha}), we employ the FNN spectral emulator with PISP-CurveFit to separately tune $n$ and $\alpha$ for both PCA and AS bases, thereby determining the default basis construction method and its optimal $(n,\alpha)$ configuration for subsequent experiments. In the second stage, with the default basis configuration and optimal $(n,\alpha)$ established, we further tune the $lr$ and $\epsilon$ of PISP-Adam on the theoretical spectrum validation set to obtain more robust optimizer settings across various inference tasks.
	
	\subsubsection{Determination of Optimal Projection Dimension $n$, Regularization Coefficient $\alpha$, and Projection Basis} \label{n-alpha}
	To determine PISP's key configuration---the basis construction method (PCA or AS) and hyperparameters $(n,\alpha)$---we first define the evaluation metric: for each setting, we compute the standard deviation of the differences ($\sigma(\Delta)$) between the inferred $25$-dimensional labels and the reference labels, and select as optimal the setting that minimizes the sum of standard deviations across all dimensions ($\sum \sigma(\Delta)$). Subsequently, under both Non-L1 and L1 settings, we compare the PCA and AS basis constructions and select the superior method as the default. To accommodate different datasets, we tune on both the Kurucz validation set and an APOGEE reference subset:
	\begin{itemize}
		\item \textbf{Hyperparameter tuning based on the Kurucz validation set.}
		\begin{itemize}
			\item \textbf{Determining $n$ for PCA-Non-L1 and AS-Non-L1 strategies.}
			 Varying $n$ from $1$ to $25$, we run inference for each setting. The subplots in row 1, column 1 (PCA-Non-L1) and row 3, column 1 (AS-Non-L1) of \reffig{valid_params} show the standard deviations of parameter differences $\sigma(\Delta)$ for different values of $n$. When $n{<}23$, the $\sigma(\Delta)$ is generally larger, indicating that basis vectors corresponding to smaller eigenvalues in either PCA or AS bases may still contain information useful for spectral fitting and thus cannot be neglected. Further comparison using the $\sum \sigma(\Delta)$ metric shows that this metric reaches its minimum at $n{=}25$. Therefore, we select $n{=}25$ as the optimal projection dimension for both PCA-Non-L1 and AS-Non-L1 strategies on Kurucz theoretical spectra.
			\item \textbf{Determining $\alpha$ for PCA-L1 and AS-L1 strategies.}
			With $n{=}25$ fixed, we conduct a grid search over $\alpha\in\{1\}\cup\{a\times 10^{-k}\mid a{=}1,{\dots},9;\,k{=}1,{\dots},4\}$. As shown in row 2, column 1 (PCA-L1) and row 4, column 1 (AS-L1) of \reffig{valid_params}, different $\alpha$ values significantly impact $\sigma(\Delta)$, with smaller $\alpha$ generally yielding better performance: for instance, under the PCA-L1 strategy, $\alpha = 10^{-3}$, $7\times10^{-4}$, and $5\times10^{-4}$ produce superior results; while under the AS-L1 strategy, $\alpha = 4\times10^{-4}$, $2\times10^{-4}$, and $3\times10^{-4}$ exhibit better performance. This indicates that excessive regularization imposes overly stringent constraints on the solution space, thereby limiting the model's ability to fit spectral details. Further comparison using the $\sum \sigma(\Delta)$ metric shows that PCA-L1 achieves optimality at $\alpha{=}7{\times}10^{-4}$, while AS-L1 is optimal at $\alpha{=}3{\times}10^{-4}$.
		\end{itemize}
		
		\item \textbf{Hyperparameter tuning based on the APOGEE reference subset.}
		 To address distributional differences between synthetic and observed spectra, we retune $n$ and $\alpha$ on APOGEE spectra. To avoid potential label leakage from the APOGEE catalog, we use the independent high-resolution analysis of \citet{Bensby2014} as a reference, which provides precise parameters for $714$ F/G dwarfs from spectra with $R{=}40{,}000$--$110{,}000$ and signal-to-noise ratio (S/N) $>150$. Cross-matching with APOGEE DR$17$ yields $45$ stars, covering $12$ of the $25$ labels considered here. We use the same hyperparameter grid as for the synthetic data. As shown in column 2 of \reffig{valid_params}, the optimal projection dimension for PCA-Non-L1 falls within $n{\in}\{7,8,9\}$, and the optimal regularization coefficient for PCA-L1 falls within $\alpha{\in}\{0.05,0.06,0.07\}$; correspondingly, the optimal $n$ for AS-Non-L1 falls within $\{6,15,25\}$, and the optimal $\alpha$ for AS-L1 falls within $\{0.02,0.03,0.05\}$. These results differ from the optimal settings for Kurucz spectra, reflecting distributional differences between theoretical and observed data; such biases can be effectively mitigated by reducing the projection dimension or increasing regularization strength. Further comparison using the $\sum \sigma(\Delta)$ metric leads to the final selection of $n{=}7$ (PCA-Non-L1) and $\alpha{=}0.06$ (PCA-L1), as well as $n{=}15$ (AS-Non-L1) and $\alpha{=}0.05$ (AS-L1) as the optimal hyperparameter settings for APOGEE.
		 
		 \item \textbf{Determining the optimal projection basis construction method.}
		 After determining the optimal projection dimensions $n$ and regularization coefficients $\alpha$ for both PCA and AS bases on theoretical and observed spectra, we use the $\sum \sigma(\Delta)$ metric to select the default projection basis construction method for subsequent experiments. As shown in \reftable{pisp_hyperparams}, PCA-Non-L1 outperforms AS-Non-L1 and PCA-L1 outperforms AS-L1 on both Kurucz theoretical spectra and the APOGEE reference subset. Therefore, in subsequent experiments (under both Non-L1 and L1 settings), we adopt PCA projection bases as the default.
	\end{itemize}
	
	\begin{figure*}[htp!]
		\centering
        \includegraphics[width=\textwidth,height=0.73\textheight, keepaspectratio]{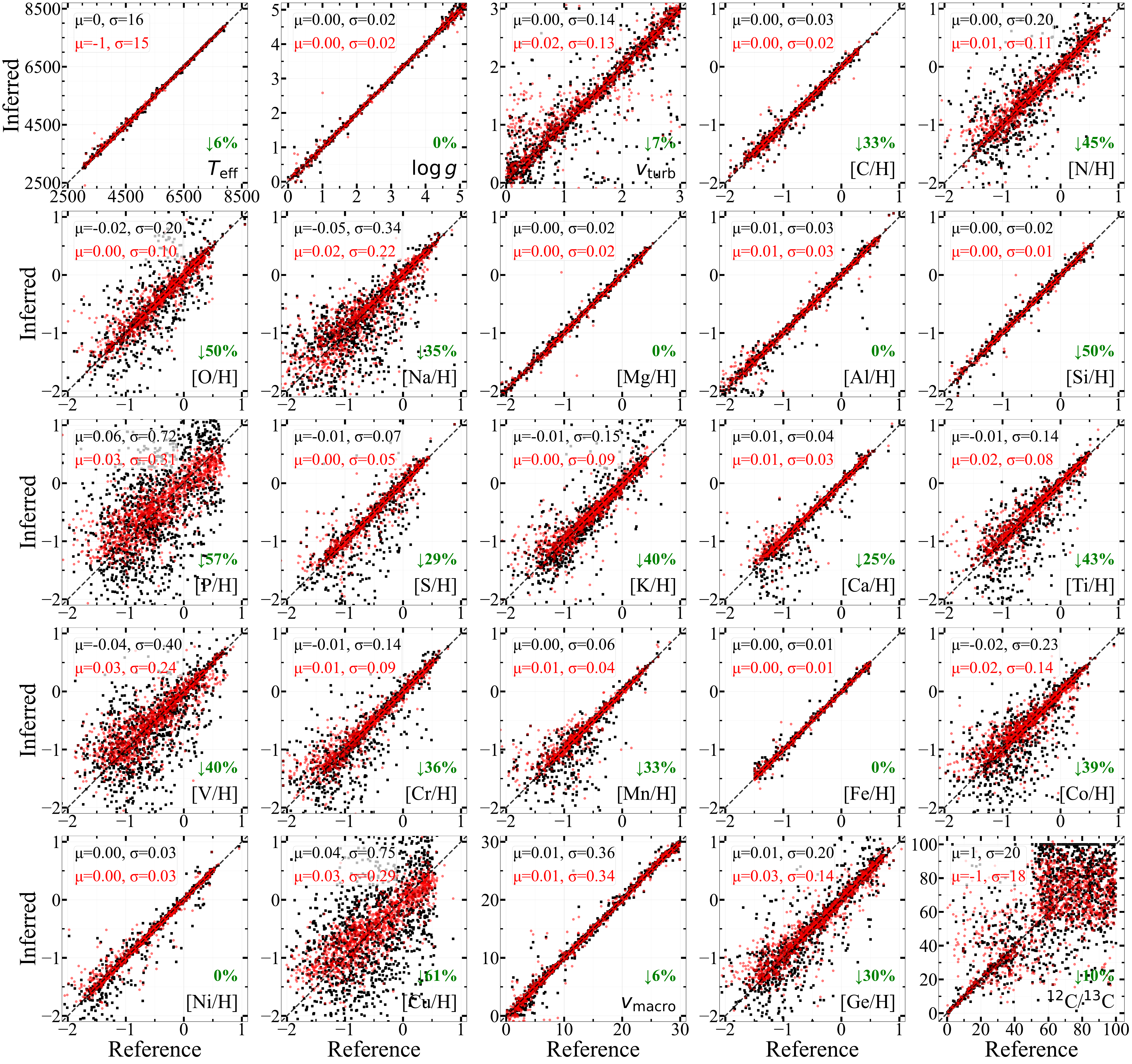}
		\caption{Comparison of stellar parameters inferred by PISP-Adam and the baseline against Kurucz reference labels on the test set. All results are based on the FNN spectral emulator and the Adam optimizer. Each panel shows inferred versus reference labels: black squares correspond to the baseline strategy, while red circles represent the PCA-L1 strategy ($\alpha{=}7{\times}10^{-4}$). Differences ($\Delta{=}\mathrm{inferred}{-}\mathrm{reference}$) statistics ($\mu$, $\sigma$) were computed via \texttt{astropy.stats.sigma\_clipped\_stats} (\texttt{sigma=3}, \texttt{maxiters=3}). Green numbers indicate the relative $\sigma$ reduction achieved by PCA-L1. The statistics are listed in the upper left in matching colors.}		
		\label{PISP_Adam_Kurucz_test_L1_vs_Raw_ANN}
	\end{figure*}
	
	\begin{figure*}[htp!]
		\centering
		\includegraphics[width=\textwidth,height=0.7\textheight, keepaspectratio]{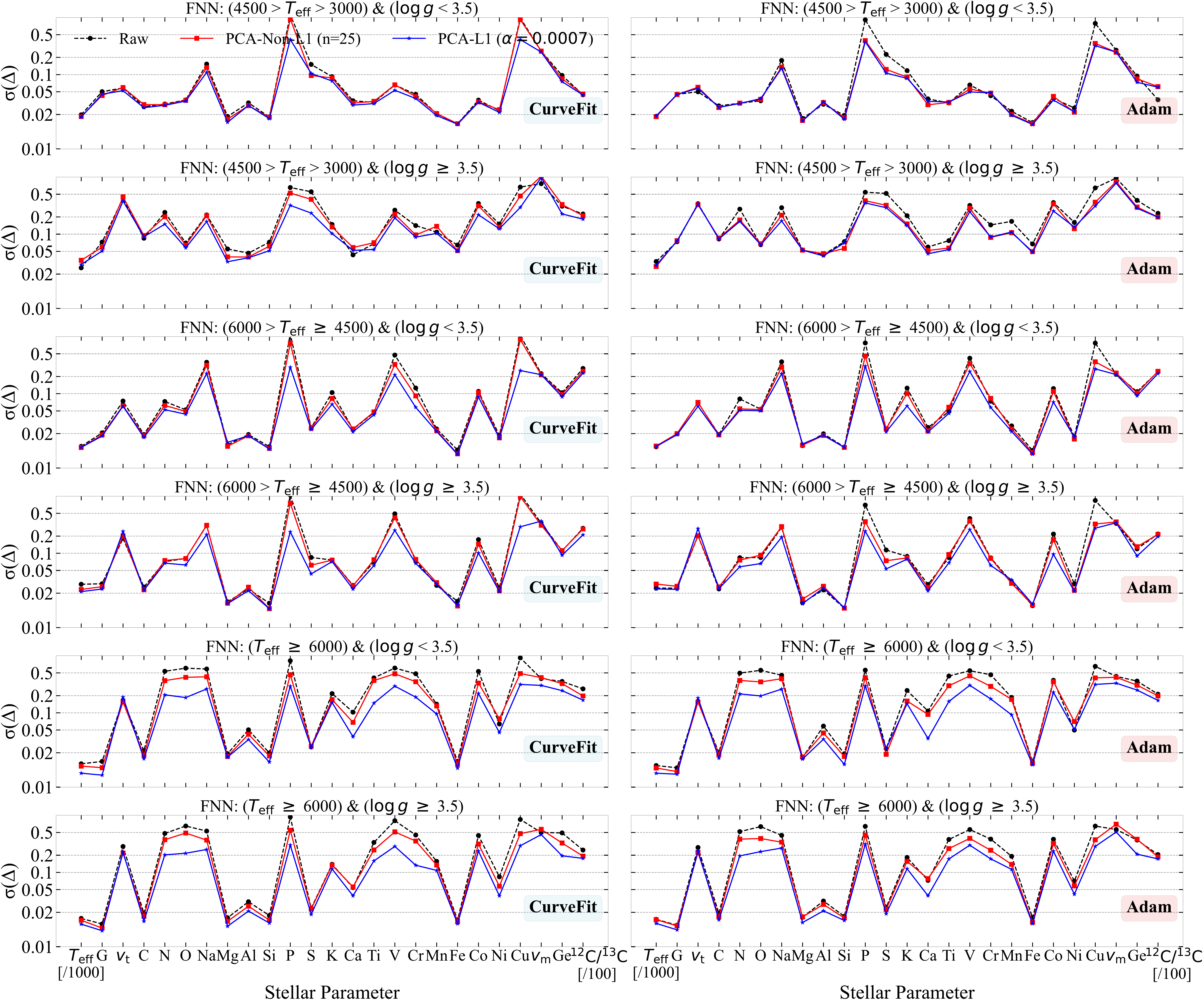}
		\caption{Standard deviations of differences relative to the Kurucz reference labels for PISP and baseline inferences on the Kurucz test set using the FNN spectral emulator. The $12$ panels are organized into six $T_{\mathrm{eff}}$--$\log g$ subsets (rows), and the optimizer changes by column (left: CurveFit; right: Adam). In each panel, $\sigma$ is computed from differences ($\Delta{=}\mathrm{inferred}{-}\mathrm{reference}$) for $25$ labels using \texttt{astropy.stats.sigma\_clipped\_stats} (\texttt{sigma=3}, \texttt{maxiters=3}). For clarity, the legend is shown only in the first panel: the black, red, and blue curves correspond to the baseline strategy described in \refsection{RawOptimization}, the PCA-Non-L1 strategy with $n=25$, and the PCA-L1 strategy with $\alpha=0.0007$, respectively, each is evaluated at its optimal hyperparameter setting. On the $x$-axis, $\log g$, $V_{\mathrm{turb}}$, and $V_{\mathrm{macro}}$ are denoted $G$, $V_t$, and $V_m$; $T_{\mathrm{eff}}$ is shown in units of $10^3\,\mathrm{K}$, and ${}^{12}\mathrm{C}/{}^{13}\mathrm{C}$ is scaled by $100$. The $y$-axis is base-10 logarithmic with tick labels given in linear units, and the dashed horizontal lines indicate reference levels to aid visual comparison across labels.}
		\label{FNN_PISP_CurveFit_vs_Adam_Kurucz_test_std}
	\end{figure*}
	
	\begin{figure*}[htp!]
		\centering
		\includegraphics[width=\textwidth,height=0.7\textheight, keepaspectratio]{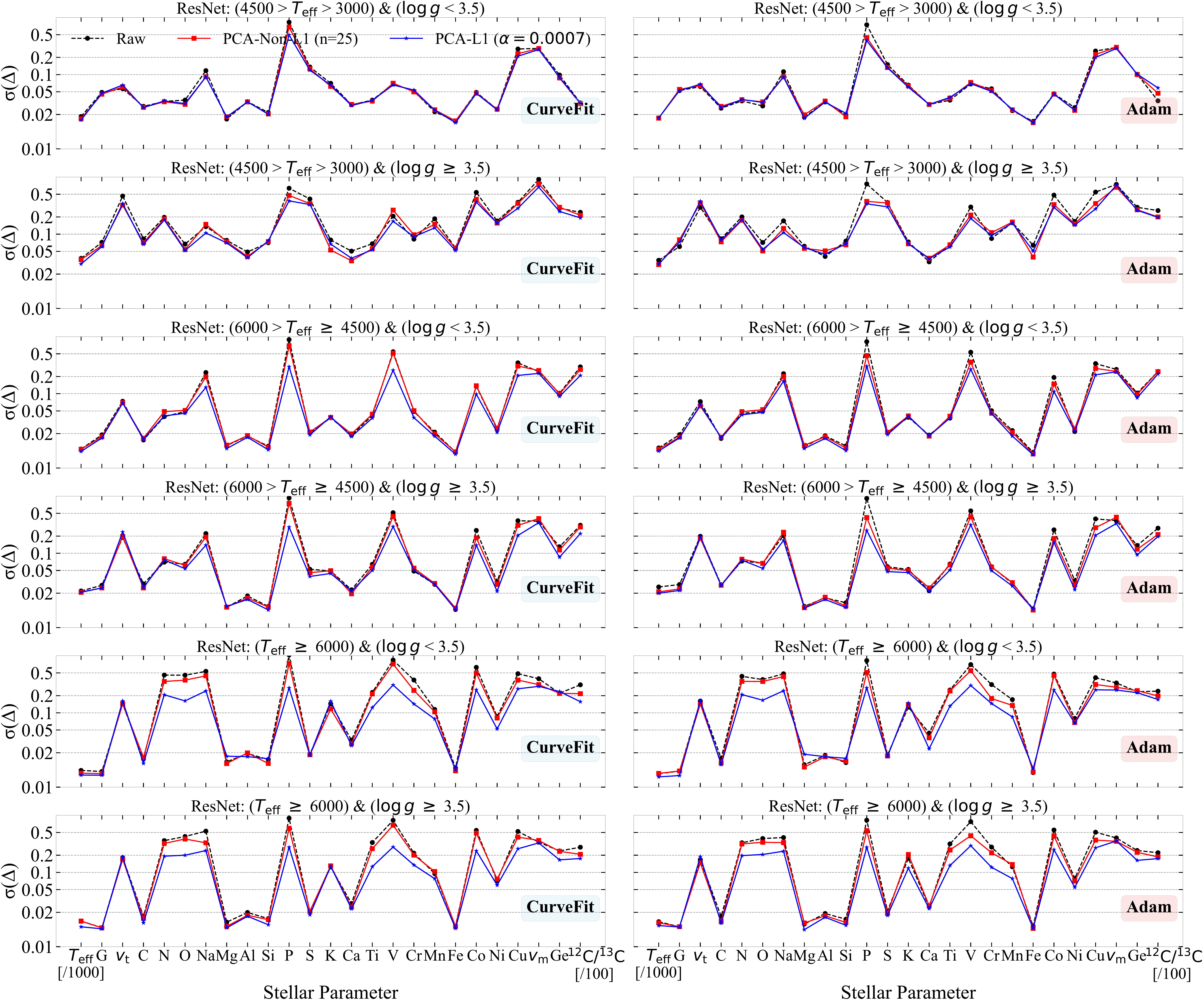}
		\caption{Standard deviations of differences relative to the Kurucz reference labels for PISP and baseline inferences on the Kurucz test set using the ResNet spectral emulator. Same as \reffig{FNN_PISP_CurveFit_vs_Adam_Kurucz_test_std}, but with the FNN emulator replaced by ResNet.}
		\label{ResNet_PISP_CurveFit_vs_Adam_Kurucz_test_std}
	\end{figure*}
	
	\begin{figure*}[htp!]
		\centering
		\includegraphics[width=\textwidth,height=0.7\textheight, keepaspectratio]{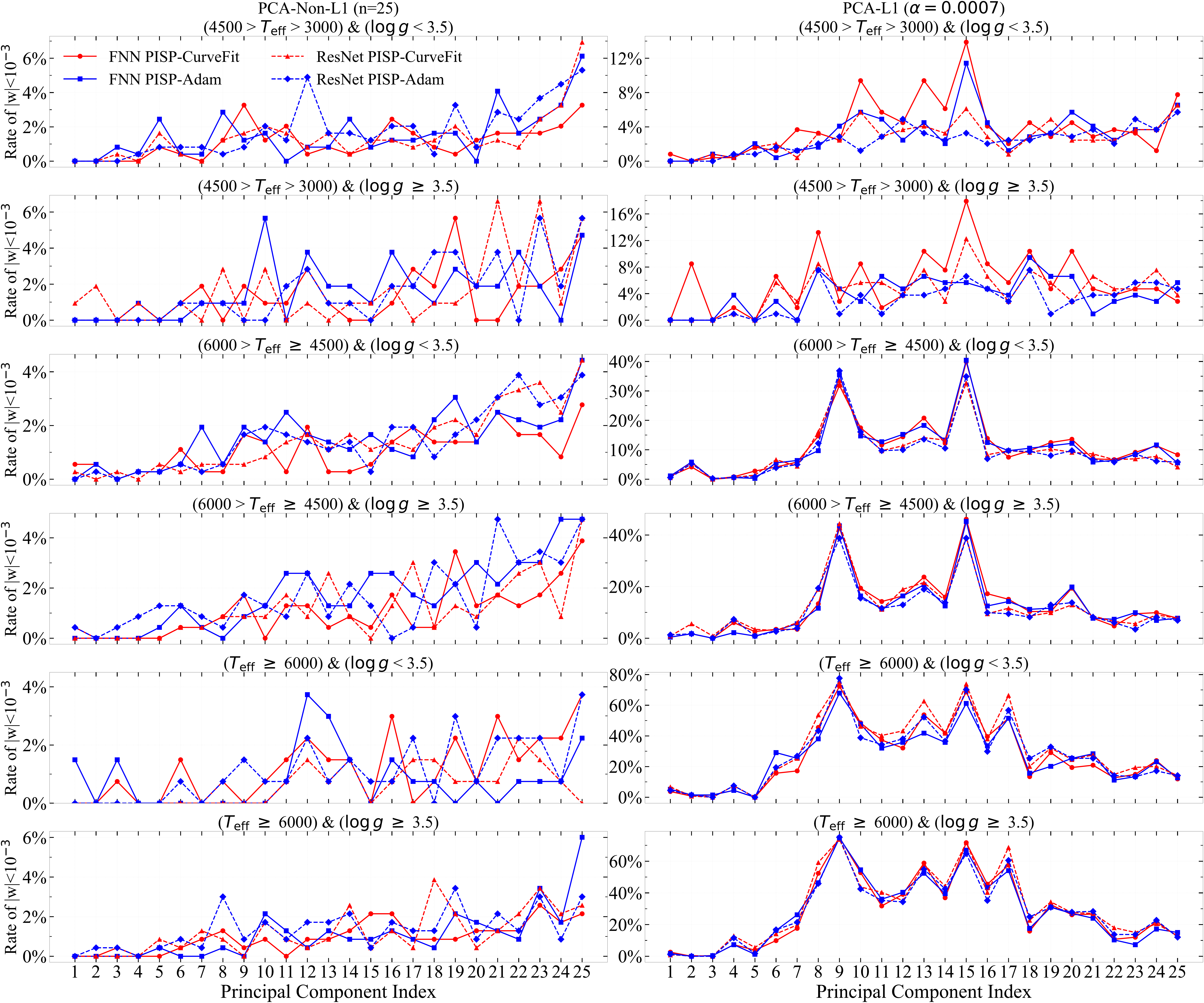}
		\caption{Fractional statistics of principal-component coefficients ($w$) with absolute values below $10^{-3}$ derived using PISP on the Kurucz test set. The first and second columns correspond to the PCA-Non-L1 and PCA-L1 strategies, respectively. The x-axis indicates the index of the principal component, and the y-axis shows the fraction of test samples with corresponding coefficients satisfying $|w| {<} 10^{-3}$. The larger the fraction, the less important the corresponding principal component is for parameter inference. Each panel represents a different $T_{\rm eff}$-$\log g$ subrange. Solid red and blue curves denote results from FNN-based spectral emulators using CurveFit and Adam, respectively, while dashed red and blue curves represent the corresponding results from ResNet-based emulators. For clarity, the legend is shown only in the first panel.}
		\label{Kurucz_PCA_vs_PCA_L1_W}
	\end{figure*}
	
	\begin{table*}[htbp]
		\centering
		\caption{Standard Deviations of Stellar Parameter Differences with Respect to APOGEE DR17 Catalog Values}
		\label{APOGEE_sigma_emulator_optimizer}
		\setlength{\tabcolsep}{7.5pt}
		\renewcommand{\arraystretch}{1.2}
		\small
		\begin{tabular}{lcccccccccccc}
			\hline\hline
			\multirow{3}{*}{\centering Parameter} &
			\multicolumn{6}{c}{FNN} &
			\multicolumn{6}{c}{ResNet} \\
			\cmidrule(lr){2-7}\cmidrule(lr){8-13}
			& \multicolumn{3}{c}{CurveFit} & \multicolumn{3}{c}{Adam}
			& \multicolumn{3}{c}{CurveFit} & \multicolumn{3}{c}{Adam} \\
			\cmidrule(lr){2-4}\cmidrule(lr){5-7}\cmidrule(lr){8-10}\cmidrule(lr){11-13}
			& Raw & Non-L1 & L1 & Raw & Non-L1 & L1 & Raw & Non-L1 & L1 & Raw & Non-L1 & L1 \\
			\hline
			$T_{\rm eff}$ & 66 & \textbf{36} & 52 & 75 & \textbf{38} & 59 & 68 & \textbf{34} & 51 & 73 & \textbf{36} & 56 \\
			$\log g$ & 0.16 & \textbf{0.13} & 0.14 & 0.17 & \textbf{0.16} & \textbf{0.16} & 0.15 & \textbf{0.13} & 0.14 & 0.16 & \textbf{0.15} & \textbf{0.15} \\
			$v_{\rm turb}$ & \textbf{0.16} & \textbf{0.16} & 0.19 & \textbf{0.20} & \textbf{0.20} & 0.21 & \textbf{0.16} & 0.17 & 0.19 & \textbf{0.18} & 0.19 & 0.20 \\
			{[C/H]} & 0.09 & 0.09 & \textbf{0.08} & 0.10 & \textbf{0.09} & \textbf{0.09} & 0.09 & 0.09 & \textbf{0.08} & 0.09 & 0.09 & \textbf{0.08} \\
			{[N/H]} & 0.09 & \textbf{0.08} & 0.09 & 0.10 & \textbf{0.09} & \textbf{0.09} & 0.09 & \textbf{0.08} & \textbf{0.08} & 0.09 & 0.09 & \textbf{0.08} \\
			{[O/H]} & 0.14 & \textbf{0.08} & 0.13 & 0.15 & \textbf{0.09} & 0.14 & 0.14 & \textbf{0.08} & 0.12 & 0.14 & \textbf{0.09} & 0.13 \\
			{[Na/H]} & 0.35 & 0.17 & \textbf{0.16} & 0.28 & \textbf{0.18} & \textbf{0.18} & 0.34 & 0.17 & \textbf{0.16} & 0.31 & \textbf{0.17} & \textbf{0.17} \\
			{[Mg/H]} & 0.08 & \textbf{0.06} & 0.09 & 0.08 & \textbf{0.07} & 0.09 & 0.08 & \textbf{0.06} & 0.09 & 0.09 & \textbf{0.06} & 0.09 \\
			{[Al/H]} & \textbf{0.11} & 0.12 & \textbf{0.11} & \textbf{0.12} & \textbf{0.12} & \textbf{0.12} & \textbf{0.11} & 0.12 & \textbf{0.11} & \textbf{0.11} & 0.12 & \textbf{0.11} \\
			{[Si/H]} & 0.11 & \textbf{0.06} & 0.12 & 0.11 & \textbf{0.07} & 0.12 & 0.11 & \textbf{0.06} & 0.12 & 0.11 & \textbf{0.07} & 0.12 \\
			{[S/H]} & 0.11 & \textbf{0.08} & 0.09 & 0.12 & \textbf{0.09} & 0.10 & 0.11 & \textbf{0.08} & 0.09 & 0.11 & \textbf{0.08} & 0.09 \\
			{[K/H]} & 0.15 & 0.13 & \textbf{0.11} & 0.16 & 0.13 & \textbf{0.12} & 0.13 & 0.13 & \textbf{0.11} & 0.14 & 0.13 & \textbf{0.11} \\
			{[Ca/H]} & \textbf{0.06} & 0.07 & \textbf{0.06} & \textbf{0.07} & \textbf{0.07} & \textbf{0.07} & \textbf{0.05} & 0.07 & 0.06 & \textbf{0.06} & 0.07 & \textbf{0.06} \\
			{[Ti/H]} & 0.13 & \textbf{0.11} & 0.12 & 0.14 & \textbf{0.11} & 0.13 & 0.13 & \textbf{0.11} & 0.12 & 0.14 & \textbf{0.11} & 0.12 \\
			{[V/H]} & 0.34 & \textbf{0.18} & 0.20 & 0.27 & \textbf{0.18} & 0.21 & 0.38 & \textbf{0.18} & 0.20 & 0.33 & \textbf{0.18} & 0.21 \\
			{[Cr/H]} & 0.11 & \textbf{0.09} & 0.11 & 0.12 & \textbf{0.09} & 0.11 & 0.10 & \textbf{0.09} & 0.10 & 0.11 & \textbf{0.09} & 0.10 \\
			{[Mn/H]} & \textbf{0.11} & \textbf{0.11} & 0.14 & 0.12 & \textbf{0.11} & 0.14 & 0.11 & \textbf{0.10} & 0.14 & 0.12 & \textbf{0.11} & 0.14 \\
			{[Fe/H]} & \textbf{0.04} & 0.05 & 0.06 & \textbf{0.05} & \textbf{0.05} & 0.07 & \textbf{0.04} & 0.05 & 0.06 & \textbf{0.05} & \textbf{0.05} & 0.06 \\
			{[Co/H]} & 0.20 & \textbf{0.16} & \textbf{0.16} & 0.20 & \textbf{0.16} & 0.17 & 0.21 & 0.16 & \textbf{0.15} & 0.22 & \textbf{0.16} & \textbf{0.16} \\
			{[Ni/H]} & \textbf{0.05} & 0.10 & \textbf{0.05} & \textbf{0.06} & 0.11 & 0.07 & \textbf{0.05} & 0.10 & \textbf{0.05} & \textbf{0.06} & 0.10 & \textbf{0.06} \\
			$v_{\rm macro}$ & 2.18 & 2.29 & \textbf{1.64} & 2.40 & 2.43 & \textbf{1.98} & 2.11 & 2.10 & \textbf{1.52} & 2.21 & 2.17 & \textbf{1.59} \\
			\hline
		\end{tabular}
		
		\vspace{0.3cm}
		\parbox{\textwidth}{\small \textbf{Note.} Values are $\sigma$ computed using \texttt{astropy.stats.sigma\_clipped\_stats} (\texttt{sigma=3}, \texttt{maxiters=3}) for differences ($\Delta{=}\mathrm{inferred}{-}\mathrm{reference}$). For brevity, `Raw' denotes the baseline strategy, `Non-L1' refers to PCA-Non-L1 ($n{=}7$), and `L1' refers to PCA-L1 ($\alpha{=}0.06$). Within each emulator--optimizer pair, the minimum $\sigma$ among the three strategies is highlighted in boldface.}
	\end{table*}
	
	\begin{figure*}[htp!]
		\centering
		\includegraphics[width=\textwidth,height=0.7\textheight, keepaspectratio]{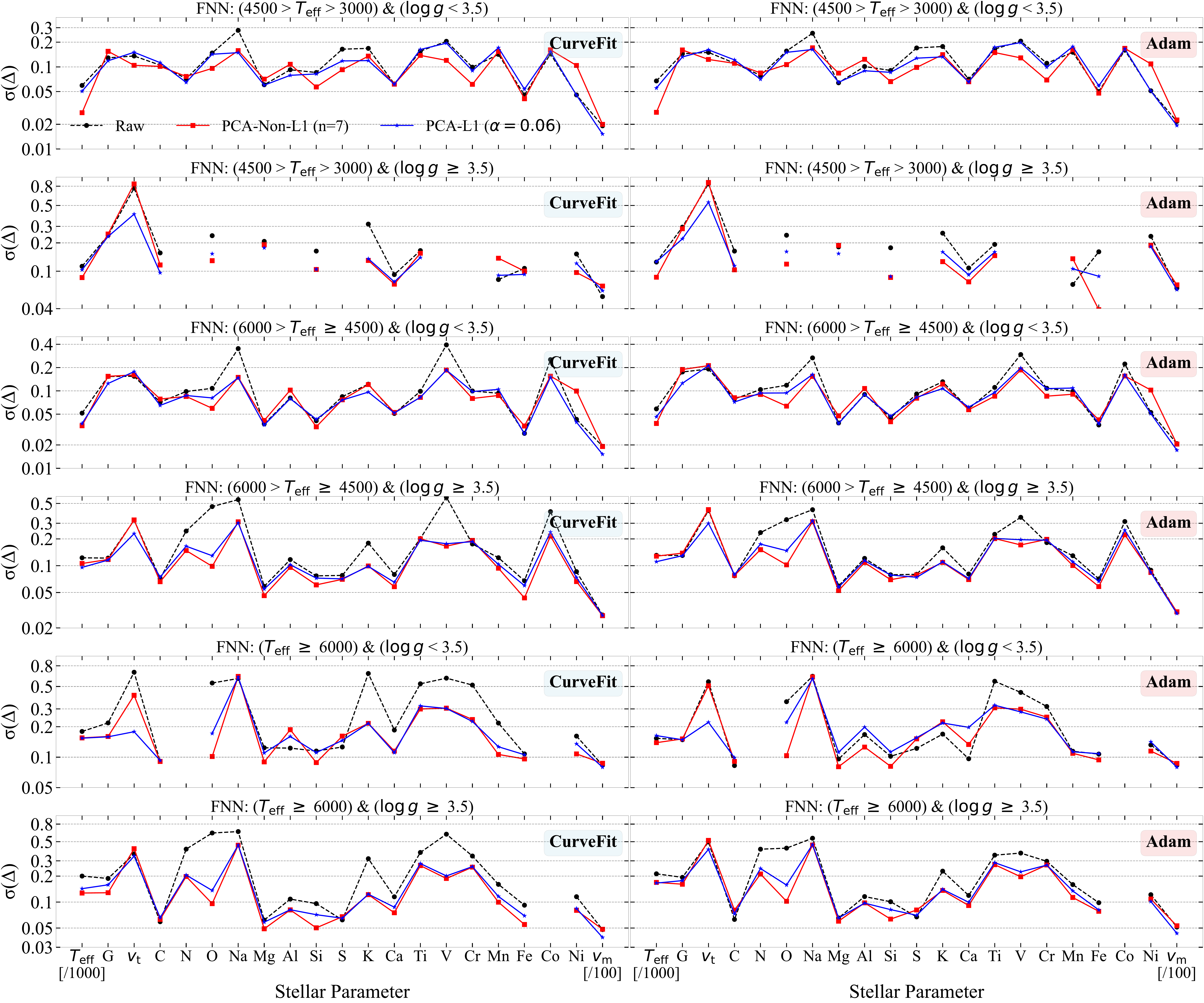}
		\caption{Standard deviations of differences relative to the APOGEE official catalog for PISP and baseline inferences on APOGEE DR17 spectra using the FNN spectral emulator. The $12$ panels are organized into six $T_{\mathrm{eff}}$--$\log g$ subsets (rows) and the optimizer changes by column (left: CurveFit; right: Adam). In each panel, $\sigma$ is computed from differences ($\Delta{=}\mathrm{inferred}{-}\mathrm{reference}$) for $21$ labels using \texttt{astropy.stats.sigma\_clipped\_stats} (\texttt{sigma=3}, \texttt{maxiters=3}). Gaps between curves indicate parameters for which APOGEE catalog values are unavailable after flag-based quality control. For clarity, the legend is shown only in the first panel: the black, red, and blue curves correspond to the baseline strategy described in \refsection{RawOptimization}, the PCA-Non-L1 strategy with $n=7$, and the PCA-L1 strategy with $\alpha=0.06$, respectively, each evaluated at its optimal hyperparameter setting. On the $x$-axis, $\log g$, $V_{\mathrm{turb}}$, and $V_{\mathrm{macro}}$ are denoted $G$, $V_t$, and $V_m$; $T_{\mathrm{eff}}$ is shown in units of $10^3\,\mathrm{K}$, and $V_m$ is scaled by $100$. The $y$-axis is base-10 logarithmic with tick labels given in linear units and the dashed horizontal lines indicate reference levels to aid visual comparison across labels.}
		\label{FNN_PISP_CurveFit_vs_Adam_APOGEE_test_std}
	\end{figure*}
	
	\begin{figure*}[htp!]
		\centering
		\includegraphics[width=\textwidth,height=0.7\textheight, keepaspectratio]{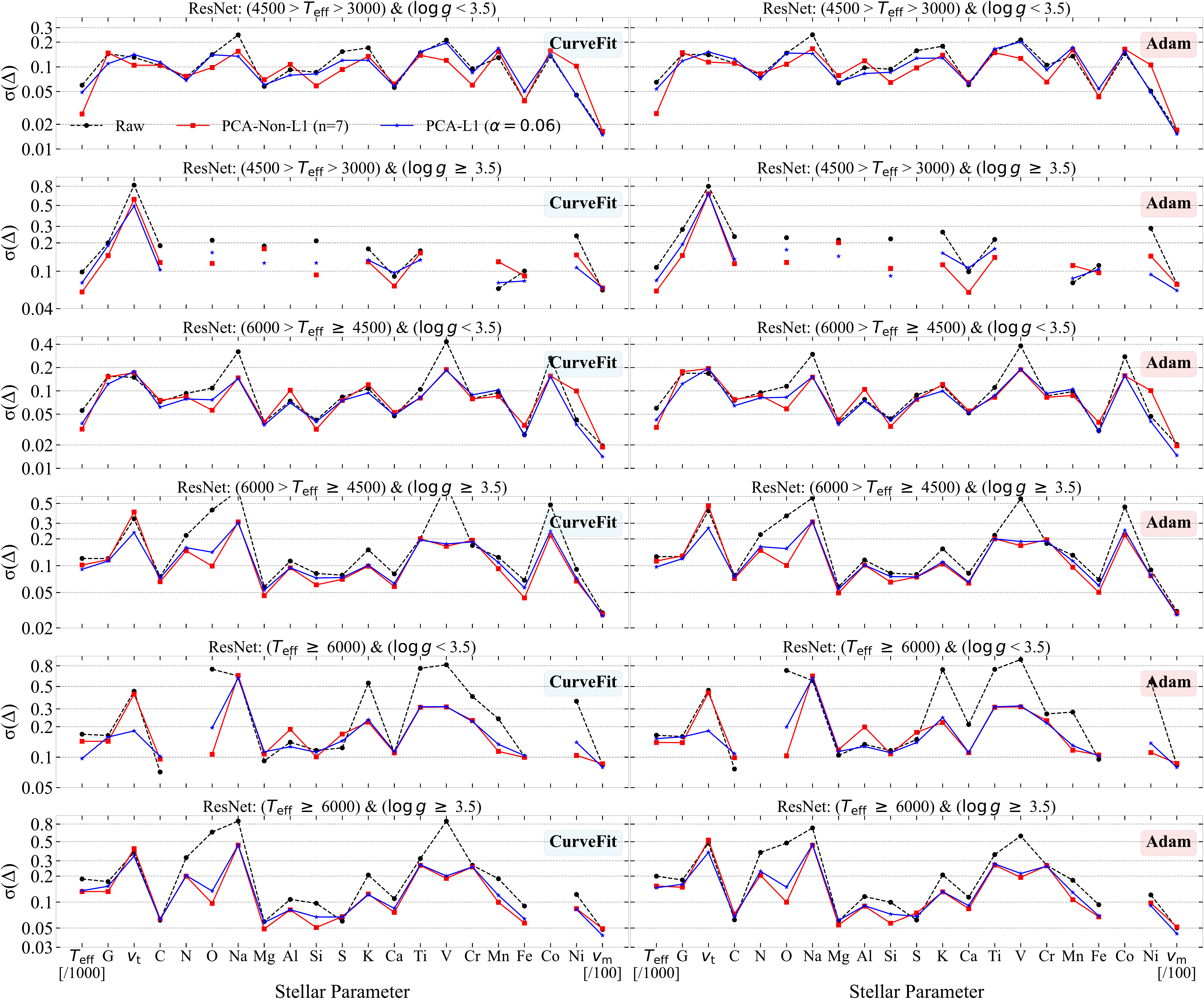}
		\caption{Standard deviations of differences relative to the APOGEE official catalog for PISP and baseline inferences on APOGEE DR17 spectra using the ResNet spectral emulator. Same as \reffig{FNN_PISP_CurveFit_vs_Adam_APOGEE_test_std}, but with the FNN emulator replaced by ResNet.}
		\label{ResNet_PISP_CurveFit_vs_Adam_APOGEE_test_std}
	\end{figure*}
	
	\begin{figure*}[htp!]
		\centering
		\includegraphics[width=\textwidth,height=0.7\textheight, keepaspectratio]{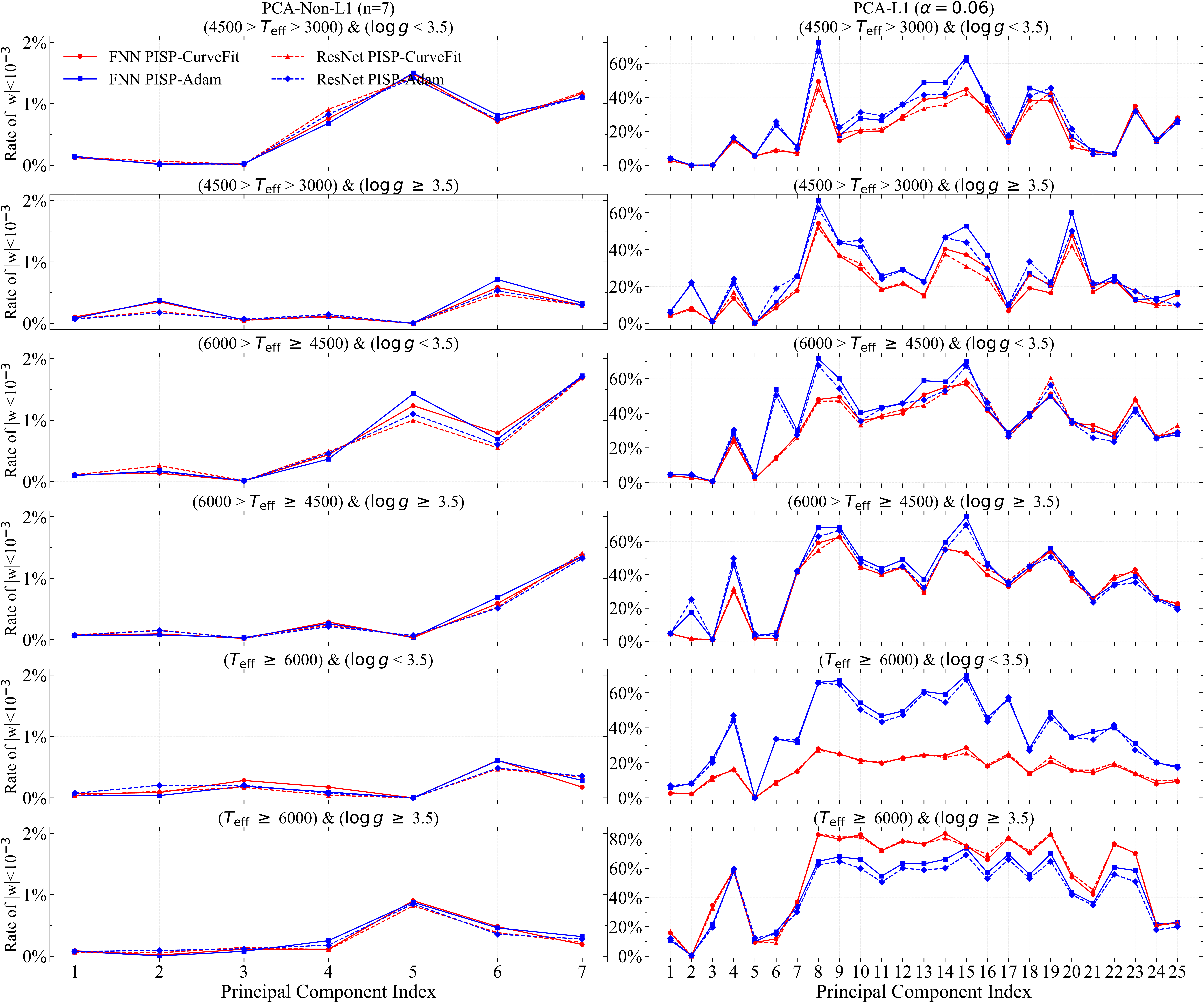}
		\caption{Fractional statistics of principal-component coefficients ($w$) with absolute values below $10^{-3}$ derived using PISP on the APOGEE spectra. The first and second columns correspond to the PCA-Non-L1 and PCA-L1 strategies, respectively. The x-axis indicates the index of the principal component and the y-axis shows the fraction of test samples with corresponding coefficients satisfying $|w| {<} 10^{-3}$. The larger the fraction, the less important the corresponding principal component is for parameter inference. For clarity, the first column shows only the first seven principal components, since the weights of the remaining $18$ principal components are $0$ (corresponding to $100\%$ on the y-axis). Each panel represents a different $T_{\rm eff}$-$\log g$ subrange. Solid red and blue curves denote results from FNN-based spectral emulators using CurveFit and Adam, respectively, while dashed red and blue curves represent the corresponding results from ResNet-based emulators. For clarity, the legend is shown only in the first panel.}
		\label{APOGEE_PCA_vs_PCA_L1_W}
	\end{figure*}
	
	\subsubsection{Learning Rate $lr$ and Convergence Threshold $\epsilon$ in PISP-Adam} \label{lr-eps}
	To determine the Adam optimizer hyperparameters $lr$ and $\epsilon$ in PISP-Adam, we fix $n{=}25$ (PCA-Non-L1) and $\alpha{=}7{\times}10^{-4}$ (PCA-L1) and evaluate the aggregate validation loss on the Kurucz validation set across different $(lr,\epsilon)$ combinations.
	
	We test three strategies: (1) the baseline of direct optimization in the standardized parameter space, (2) PCA-Non-L1, and (3) PCA-L1. For each strategy, we set $\epsilon\in\{10^{-i}\mid i{=}4,5,6,7,8\}$; for each $\epsilon$ we scan $lr\in\{1\}\cup\{a{\times}10^{-k}{\mid} a{=}1,{\dots},9;\,k{=}1,{\dots},5\}$. \reffig{Adam_best_lr} shows the distribution of aggregate validation loss over $(lr,\epsilon)$. The results indicate that, for a fixed $lr$, smaller $\epsilon$ generally yields lower loss, with the minimum at $\epsilon{=}10^{-8}$; all three strategies exhibit good convergence for $lr{\in}[0.001,\,0.01]$, suggesting this is a robust range. Furthermore, at $\epsilon{=}10^{-8}$, the runtime with $lr{=}0.001$ is approximately $6\times$, $7\times$, and $6\times$ that with $lr{=}0.01$ for the baseline, PCA-Non-L1, and PCA-L1, respectively, indicating that $lr{=}0.01$ substantially reduces inference cost while maintaining low loss. Balancing convergence and efficiency, we recommend $lr{=}0.01$ and $\epsilon{=}10^{-8}$ as default settings for PISP-Adam to support stable and efficient large-scale inference.
	
	\subsection{Parameter Inference on Synthetic Spectra} \label{inference-synthetic}
	To evaluate the performance of PISP-CurveFit and PISP-Adam on the theoretical spectrum test set, we adopt the optimal PISP configuration determined in \refsubsection{pisp-hyperopt}---PCA projection bases with hyperparameters ($n{=}25$, $\alpha{=}7 {\times} 10^{-4}$, $lr{=}0.01$, $\epsilon{=}10^{-8}$)---and compare the two PCA-assisted strategies (PCA-Non-L1 and PCA-L1) with the baseline strategy.
		
	\reftable{Kurucz_sigma_emulator_optimizer} presents the standard deviation of parameter differences $\sigma(\Delta)$ for two spectral emulators and two optimizers under three inference strategies (Raw, PCA-Non-L1, PCA-L1). Across most label dimensions, PCA-L1 performs best: $12$ of $20$ elemental abundances show $\sigma(\Delta)$ reductions of at least $0.01$ dex relative to the baseline strategy in all four emulator-optimizer combinations, with [N/H], [O/H], [Na/H], [Co/H], [P/H], [V/H], and [Cu/H] showing reductions of $0.05$--$0.72$ dex. Consider the case of FNN+Adam: PCA-L1 yields notably lower $\sigma(\Delta)$ than the baseline strategy for multiple elemental abundances---[P/H] decreases from $0.72$ to $0.31$ dex, [Cu/H] from $0.75$ to $0.29$ dex, [V/H] from $0.40$ to $0.24$ dex, while [N/H] and [O/H] decrease from $0.20$ dex to $0.11$ dex and $0.10$ dex, respectively. For parameters with low baseline $\sigma(\Delta)$ (e.g., $\log g$ and [Fe/H]), differences among strategies are minimal. \reffig{PISP_Adam_Kurucz_test_L1_vs_Raw_ANN} illustrates this comparison for FNN+Adam, showing inferred versus reference labels under both the baseline and PCA-L1 strategies. Relative to the baseline strategy, PCA-L1 reduces $\sigma(\Delta)$ by $25\%{-}61\%$ for abundance labels, and systematic biases show minimal variation. Similar trends emerge across other emulator-optimizer combinations, demonstrating consistent improvements.
	
	We further evaluate PISP performance across six $T_{\mathrm{eff}}$--$\log g$ subregions (\reffig{FNN_PISP_CurveFit_vs_Adam_Kurucz_test_std} and \reffig{ResNet_PISP_CurveFit_vs_Adam_Kurucz_test_std}). For both FNN and ResNet emulators, PISP reduces $\sigma(\Delta)$ for most elemental abundances relative to the baseline strategy. Consider PCA-L1 with FNN+Adam (\reffig{FNN_PISP_CurveFit_vs_Adam_Kurucz_test_std}): defining $\Delta\sigma {=} \sigma_{\rm Raw}{-}\sigma_{\rm PCA\text{-}L1}$, we adopt $\Delta\sigma {\ge} 0.01$ dex as the criterion for \lq effective improvement\rq \ in elemental abundances. In the cool temperature regime ($3000 {<} T_{\mathrm{eff}} {<} 4500$), $7$ ($\log g {<} 3.5$) and $15$ ($\log g {\ge} 3.5$) elemental abundances satisfy this criterion; in the intermediate temperature regime ($4500 {\le} T_{\mathrm{eff}} {<} 6000$), the corresponding numbers are $9$ ($\log g {<} 3.5$) and $11$ ($\log g {\ge} 3.5$); in the hot temperature regime ($T_{\mathrm{eff}} {\ge} 6000$), the numbers increase to $14$ ($\log g {<} 3.5$) and $15$ ($\log g {\ge} 3.5$). Elements showing \lq effective improvement\rq\ across all six subregions include [Na/H], [P/H], [Cu/H], [V/H], and [Ge/H], with $\Delta\sigma$ ranges of $0.05$--$0.19$ dex, $0.19$--$0.53$ dex, $0.33$--$0.57$ dex, $0.02$--$0.26$ dex, and $0.02$--$0.16$ dex, respectively. Additionally, \lq effective improvements\rq\ for certain elements exhibit temperature dependence: for instance, improvements for [N/H] and [O/H] primarily occur in the hot temperature regime, with $\Delta\sigma_{\mathrm{[N/H]}}{=}0.28$--$0.32$ dex and $\Delta\sigma_{\mathrm{[O/H]}}{=}0.36$--$0.40$ dex; while [Ti/H] and [Cr/H] also achieve $\Delta\sigma_{\mathrm{[Ti/H]}}{=}0.21$--$0.28$ dex and $\Delta\sigma_{\mathrm{[Cr/H]}}{=}0.21$--$0.29$ dex in the hot regime. In contrast, for parameters such as $T_{\mathrm{eff}}$ and $\log g$, the improvements in $\sigma$ are on a smaller scale (e.g., $\Delta\sigma_{T_{\mathrm{eff}}} {\le} 5$ K, $\Delta\sigma_{\log g} {\le} 0.002$ dex). In a few cases, the baseline strategy performs better, such as in the subregion with $3000 {<} T_{\mathrm{eff}} {<} 4500$ and $\log g {<} 3.5$, where [O/H], [Al/H], [Ti/H], and [Cr/H] correspond to $\Delta\sigma$ values of $-0.003$, $-0.003$, $-0.001$, and $-0.005$ dex, respectively. These results demonstrate that PISP primarily enhances precision for elemental abundances, with consistent improvements across temperature and gravity regimes, though isolated cases exhibit $\Delta\sigma {\le} 0$.
	
	Notably, although PCA-Non-L1 improves inference accuracy over the baseline strategy, it still underperforms PCA-L1 for most labels (see \reftable{Kurucz_sigma_emulator_optimizer}, or the red and blue solid lines in \reffig{FNN_PISP_CurveFit_vs_Adam_Kurucz_test_std} and \reffig{ResNet_PISP_CurveFit_vs_Adam_Kurucz_test_std}). Comparing weight distributions across basis vectors (\reffig{Kurucz_PCA_vs_PCA_L1_W}) reveals the cause: PCA-Non-L1 assigns lower weights to low-variance basis vectors (higher fraction of $|w|{<}10^{-3}$), whereas PCA-L1 concentrates weights on the first and last seven basis vectors while suppressing intermediate ones. This indicates that when training and test distributions are consistent (both from Kurucz synthetic spectra), low-variance basis vectors carry information crucial for parameter inference, and PCA-L1's ability to adaptively select relevant basis vectors yields superior performance compared to PCA-Non-L1. We therefore recommend PCA-L1 as the default for matched distributions.
	
	\subsection{Parameter Inference on Observed Spectra} \label{pisp-observed}
	To assess PISP's applicability to real observational data, we apply PISP-CurveFit and PISP-Adam to APOGEE DR$17$ observed spectra, adopting the optimal PISP configuration determined in \refsubsection{pisp-hyperopt}---PCA projection bases with hyperparameters: $n{=}7$, $\alpha{=}0.06$, $lr{=}0.01$, $\epsilon{=}10^{-8}$. We compare against APOGEE official catalog labels, selecting samples with $T_{\mathrm{eff}}{\in}[3000, 8000]\,\mathrm{K}$ and \texttt{STAR\_FLAG}, \texttt{RV\_FLAG}, \texttt{ASPCAP\_FLAG}, as well as individual parameter \texttt{FLAG} values, all equal to zero.
	
	\reftable{APOGEE_sigma_emulator_optimizer} presents the standard deviations of parameter differences $\sigma(\Delta)$ relative to APOGEE official catalog labels for two spectral emulators and two optimizers under three inference strategies (Raw, PCA-Non-L1, PCA-L1). Across most stellar-parameter dimensions, PCA-Non-L1 and PCA-L1 achieve the smallest $\sigma(\Delta)$, outperforming the baseline strategy: PCA-Non-L1 reduces $\sigma(\Delta)$ by more than $30$ K for $T_{\mathrm{eff}}$ and by at least $0.01$ dex for $9$ of $17$ elemental abundances in all four emulator-optimizer combinations, with [O/H], [Na/H], [V/H] showing reductions of $0.05$--$0.20$ dex. For example, with ResNet+CurveFit, PCA-Non-L1 reduces $T_{\mathrm{eff}}$ from $68$ to $34$ K, [O/H] from $0.14$ to $0.08$ dex, [Na/H] from $0.34$ to $0.17$ dex, [Si/H] from $0.11$ to $0.06$ dex, [V/H] from $0.38$ to $0.18$ dex, and [Co/H] from $0.21$ to $0.16$ dex. However, the improvements are not uniform: for [Al/H], [Ca/H], and [Fe/H], the baseline strategy retains a $\sigma(\Delta)$ advantage of approximately $0.01$--$0.05$ dex over PCA-Non-L1.
	
	\reffig{FNN_PISP_CurveFit_vs_Adam_APOGEE_test_std} and \reffig{ResNet_PISP_CurveFit_vs_Adam_APOGEE_test_std} compare $\sigma(\Delta)$ between PISP and the baseline strategy across six $T_{\mathrm{eff}}$--$\log g$ subregions. Due to FLAG restrictions in the APOGEE official catalog, the number of available reference parameters varies across subregions, ranging from $14$ to $21$ parameters. Overall, both PCA-L1 and PCA-Non-L1 outperform the baseline strategy for both emulators, though their relative advantages vary across $T_{\mathrm{eff}}$--$\log g$ subregions. For instance, in the giant regime at $4500 {\le} T_{\mathrm{eff}} {<} 6000$, PCA-L1 achieves lower $\sigma(\Delta)$ for $13$ parameters while PCA-Non-L1 achieves lower $\sigma(\Delta)$ for six parameters; in the dwarf regime at $4500 {\le} T_{\mathrm{eff}} {<} 6000$, PCA-Non-L1 outperforms PCA-L1 for $14$ parameters versus six; in the dwarf regime at $T_{\mathrm{eff}} {\ge} 6000$, PCA-Non-L1 leads for $14$ parameters versus four for PCA-L1, while the remaining subregions show comparable performance. Combining the global statistics from \reftable{APOGEE_sigma_emulator_optimizer} with the $T_{\mathrm{eff}}$--$\log g$ subregion analysis reveals that PCA-Non-L1 slightly outperforms PCA-L1 in the number of parameters achieving superior performance on APOGEE observed spectra, contrasting with PCA-L1's dominance on the theoretical spectrum test set. To understand this discrepancy, we compare the weight distributions across basis vectors for both strategies. As shown in \reffig{APOGEE_PCA_vs_PCA_L1_W}, PCA-L1 and PCA-Non-L1 exhibit different weight allocation patterns across basis vectors. Based on this observation, we offer the following qualitative interpretation: spectral features represented by low-variance basis vectors may be more susceptible to noise and model-observation mismatches in observed data. Consequently, PCA-L1's stronger emphasis on these low-variance basis vectors may limit its precision gains, while PCA-Non-L1's focus on high-variance basis vectors may be more robust. In practice, strategy selection should be guided by data quality and the consistency between training and target distributions: when spectral noise levels are high or significant domain shifts exist, PCA-Non-L1 may be preferable; when data quality is high and distributions are well-matched, PCA-L1 may still provide additional precision improvements.
	
	\begin{figure*}[htp!]
		\centering
		\includegraphics[width=\textwidth, height=0.7\textheight, keepaspectratio]{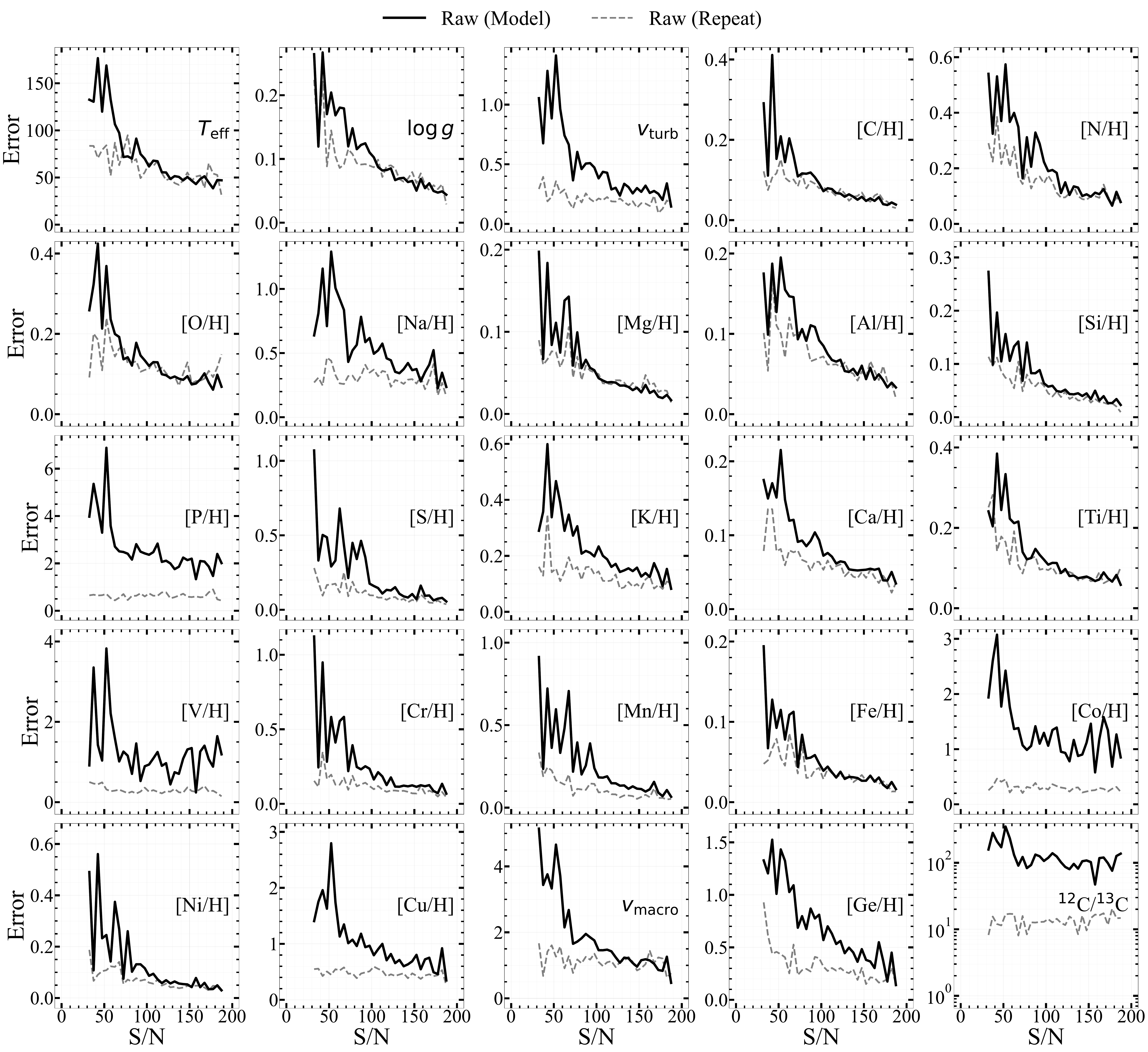}
		\caption{Comparison between model errors (solid curves) and repeat-observation errors (dashed curves) as a function of S/N for the baseline strategy. All results are based on the ResNet spectral emulator and PISP-CurveFit. Errors are computed in S/N bins of width $5$ and are reported only for bins containing at least three stars, with the median error shown on the vertical axis. For $^{12}\mathrm{C}/^{13}\mathrm{C}$, the $y$-axis is base-10 logarithmic with tick labels given in linear units.}
		\label{ResNet_Raw_CurveFit_repeat_err_SNR}
	\end{figure*}
	
	\begin{figure*}[htp!]
		\centering
		\includegraphics[width=\textwidth, height=0.7\textheight, keepaspectratio]{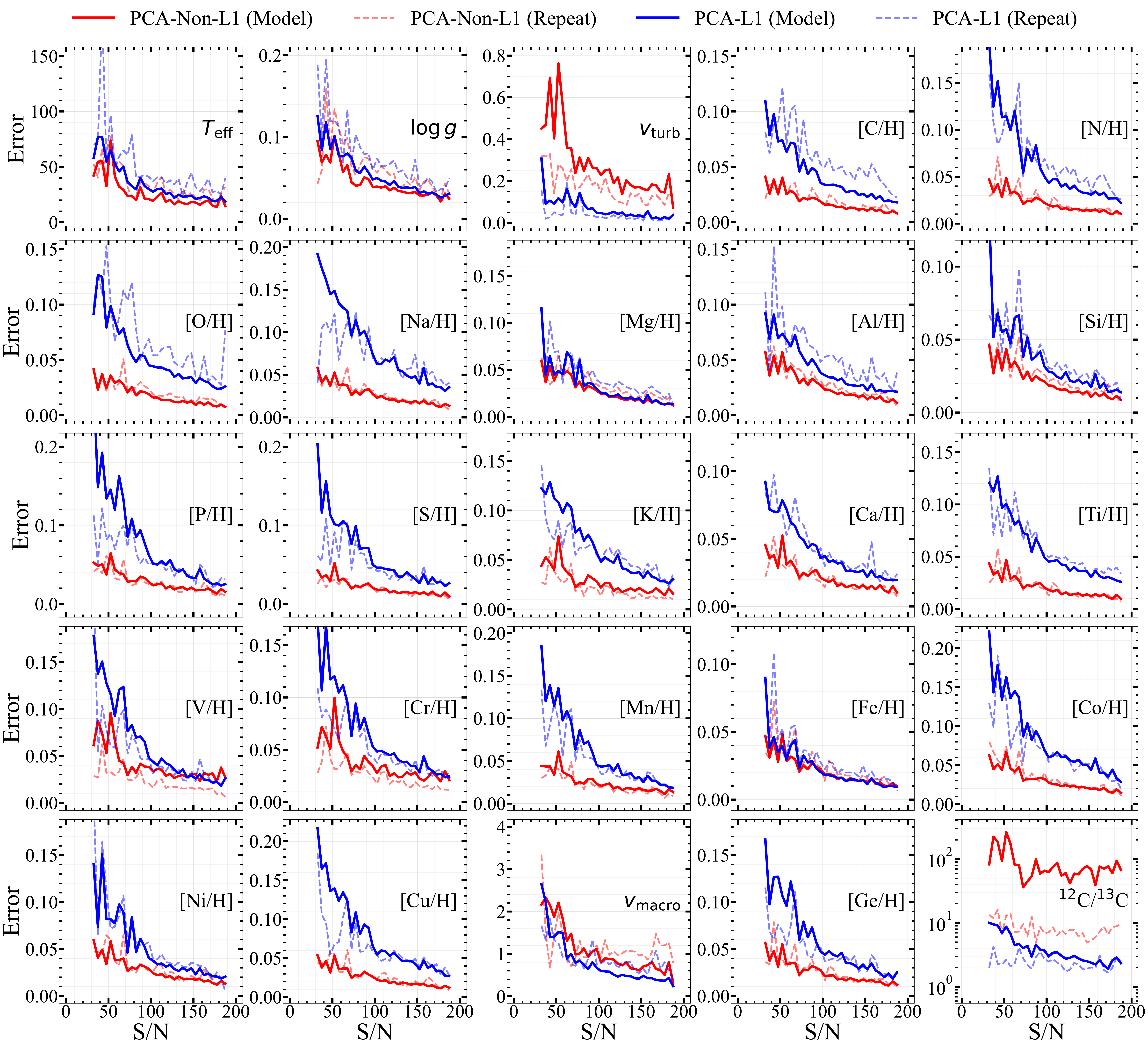}
		\caption{Comparison between model errors (solid curves) and repeat-observation errors (dashed curves) as a function of S/N for PISP. Same as \reffig{ResNet_Raw_CurveFit_repeat_err_SNR}, except that the curves are color-coded, with red for  PCA-Non-L1 ($n{=}7$) and blue for PCA-L1 ($\alpha{=}0.06$).}
		\label{ResNet_PISP_CurveFit_repeat_err_SNR}
	\end{figure*}
	
	\subsection{Error Analysis} \label{Error}
	To evaluate the validity of PISP's model errors, we compare model errors with repeat-observation errors, using the inference results from the ResNet emulator combined with PISP-CurveFit as an example. Model errors are derived via first-order linear error propagation in the neighborhood of the optimal solution, estimating parameter errors arising from the propagation of spectral noise from pixel space to parameter space. Repeat-observation errors originate from the dispersion of parameters across multiple independent observations of the same target, reflecting the combined effects of noise and systematic variations. Since all spectra employ consistent normalization and pixel masking strategies (\refsection{Data Preprocessing}), differences introduced by continuum processing and outlier pixels are minimized; thus, repeat-observation errors serve as an empirical reference for model errors. If model errors are reasonable, the two should be statistically consistent and exhibit similar S/N dependencies. For this analysis, we select $5939$ spectra from $1049$ stars out of $722{,}896$ APOGEE spectra, where each star has at least four independent observations with S/N not exceeding $200$.
	
	As shown in \reffig{ResNet_Raw_CurveFit_repeat_err_SNR}, under the baseline strategy, both model errors and repeat-observation errors generally decrease with increasing S/N. However, when S/N falls below approximately $70$, the two diverge significantly: model errors for most parameters are systematically overestimated, such as for $T_{\mathrm{eff}}$, $\log g$, and multiple elemental abundances, where model errors in the low-S/N regime can reach approximately twice the repeat-observation errors. Additionally, certain parameters exhibit extreme error estimates, such as [P/H], [V/H], [Co/H], and [Cu/H], where model errors far exceed repeat-observation errors. This indicates that for spectral emulators, error estimates based on first-order linearization in the neighborhood of the optimal solution may not be sufficiently accurate; particularly when the model exhibits strong nonlinearity with respect to certain parameters, this linearization approximation can introduce biases \citep{Vugrin2007}. A detailed mechanistic analysis of this phenomenon is beyond the scope of this study. In contrast, the PISP strategy improves this behavior for all parameters except $^{12}\mathrm{C}/^{13}\mathrm{C}$ (see \reffig{ResNet_PISP_CurveFit_repeat_err_SNR}): model errors no longer show the extreme discrepancies observed in the baseline strategy (e.g., for [P/H], [V/H], [Co/H], and [Cu/H]), and exhibit stronger consistency with repeat-observation errors in the S/N ${<}70$ regime; particularly for S/N ${>}70$, the magnitude and trends of model errors and repeat-observation errors are largely consistent. This demonstrates that while improving parameter-inference precision, PISP also enhances the empirical consistency between model errors and repeat-observation errors.
	
	\begin{figure*}[htp!]
		\centering
		\includegraphics[width=\textwidth,height=0.7\textheight, keepaspectratio]{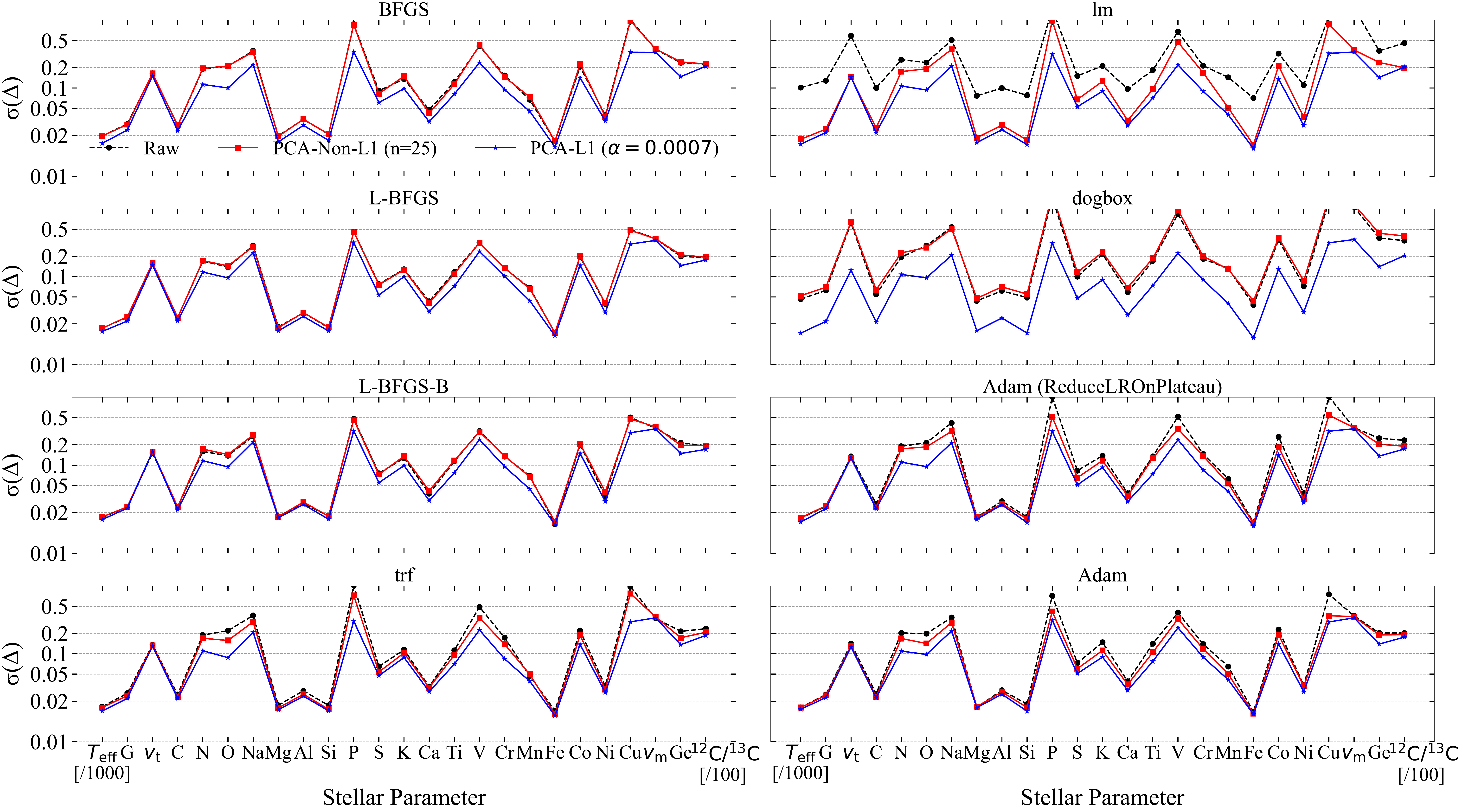}
		\caption{Standard deviations of differences relative to the Kurucz reference labels for PISP and baseline inferences on the test set obtained with different optimizers using the FNN spectral emulator. Each panel corresponds to a different optimizer, including BFGS, L-BFGS, L-BFGS-B, trf, lm, dogbox, Adam with ReduceLROnPlateau, and Adam. In each panel, $\sigma$ is computed from differences ($\Delta{=}\mathrm{inferred}{-}\mathrm{reference}$) for $25$ labels using \texttt{astropy.stats.sigma\_clipped\_stats} (\texttt{sigma=3}, \texttt{maxiters=3}). For clarity, the legend is shown only in the first panel: the black, red, and blue curves correspond to the baseline strategy described in \refsection{RawOptimization}, the PCA-Non-L1 strategy with $n=25$, and the PCA-L1 strategy with $\alpha=0.0007$, respectively, each evaluated at its optimal hyperparameter setting. On the $x$-axis, $\log g$, $V_{\mathrm{turb}}$, and $V_{\mathrm{macro}}$ are denoted $G$, $V_t$, and $V_m$; $T_{\mathrm{eff}}$ is shown in units of $10^3\,\mathrm{K}$, and ${}^{12}\mathrm{C}/{}^{13}\mathrm{C}$ is scaled by $100$. The $y$-axis is base-10 logarithmic with tick labels given in linear units and the dashed horizontal lines indicate reference levels to aid visual comparison across labels.}
		\label{PISP_different_optimizers_comparison}
	\end{figure*}
	
	\subsection{Optimizer Dependence under Orthogonal Reparameterization} \label{different-optimizers}
	To systematically assess whether PISP's performance improvement depends on the choice of optimizer, we compare parameter inference results using the FNN emulator across multiple commonly used optimizers on the theoretical spectrum test set. Beyond the default trf \citep{Coleman1996,Branch1999} and Adam \citep{Kingma2014} employed in this work, we further test Broyden--Fletcher--Goldfarb--Shanno (BFGS; \citeauthor{Nocedal2006} \citeyear{Nocedal2006}), limited-memory BFGS (L-BFGS; \citeauthor{Nocedal2006} \citeyear{Nocedal2006}), Levenberg--Marquardt (lm; \citeauthor{More1978} \citeyear{More1978}), limited-memory BFGS with bound constraints (L-BFGS-B; \citeauthor{Byrd1995} \citeyear{Byrd1995}; \citeauthor{Zhu1997} \citeyear{Zhu1997}), dogleg algorithm with rectangular trust regions (dogbox; \citeauthor{Voglis2004} \citeyear{Voglis2004}; \citeauthor{Nocedal2006} \citeyear{Nocedal2006}), and Adam with ReduceLROnPlateau. Under ideal conditions where initialization, line search, and stopping criteria are identical and numerical errors are negligible, BFGS and L-BFGS can be regarded as unconstrained methods insensitive to orthogonal reparameterization (hereafter referred to as rotation invariance). Therefore, under the setting of orthogonal reparameterization of standardized stellar parameters using \refformula{PISP-CurveFit-PCA-2} without dimensionality reduction (i.e., PCA-Non-L1 with $n{=}25$), their results should theoretically remain essentially identical to the baseline strategy; for these optimizers, we primarily compare $\sigma(\Delta)$ between PCA-L1 ($\alpha{=}7\times10^{-4}$) and the baseline strategy. In contrast, L-BFGS-B, lm, dogbox, trf, and Adam are typically more sensitive to orthogonal reparameterization (i.e., not rotation invariant) due to mechanisms such as bound constraints, trust regions, damping strategies, or per-coordinate adaptive scaling; for these optimizers, we compare both PCA-Non-L1 ($n{=}25$) and PCA-L1 ($\alpha{=}7\times10^{-4}$) against the baseline strategy to characterize the impact of both approaches. Throughout this work, we use the term rotation invariance to refer strictly to the theoretical property under the idealized conditions described in this section; any observed differences under practical implementations should be interpreted as resulting from optimizer-specific mechanisms or numerical errors, rather than a violation of the theoretical property itself.
	
	\begin{table*}[htp!]
		\centering
		\caption{Inference Time on APOGEE DR17 Spectra using CurveFit and Adam with Baseline, PCA-Non-L1, and PCA-L1 Strategies}
		\label{APOGEE_Time}
		\renewcommand{\arraystretch}{1}
		\begin{tabular*}{\textwidth}{@{\extracolsep{\fill}}lcccccc@{}}
			\toprule
			\multirow{2}{*}{Method} & \multicolumn{3}{c}{FNN} & \multicolumn{3}{c}{ResNet} \\
			\cmidrule{2-4} \cmidrule{5-7}
			& baseline & PCA-Non-L1 & PCA-L1 & baseline & PCA-Non-L1 & PCA-L1 \\
			\midrule
			CurveFit & 20.78 h & 4.99 h & 22.45 h & 27.62 h & 6.59 h & 29.51 h \\
			\addlinespace[3pt]        
			Adam & 0.50 h & 0.38 h & 0.49 h & 8.37 h & 10.33 h & 7.87 h \\
			\bottomrule
		\end{tabular*}
	\end{table*}
	
	As shown in \reffig{PISP_different_optimizers_comparison}, PISP's performance across different optimizer settings can be summarized in three key observations:
	\begin{itemize}
		\item \textbf{For rotation-invariant optimizers, PCA-Non-L1 ($n{=}25$) is essentially equivalent to the baseline strategy.}
		For BFGS and L-BFGS, PCA-Non-L1 ($n{=}25$) yields $\sigma(\Delta)$ closest to that of the baseline strategy, indicating that without dimensionality reduction or L1 regularization, orthogonal reparameterization alone has limited impact on these optimizers. This is because under the transformation defined in \refformula{PISP-CurveFit-PCA-2}, the optimization problems corresponding to \refformula{PISP-CurveFit-PCA-loss1} and \refformula{PISP-CurveFit-PCA-loss2} are mathematically equivalent: under the aforementioned ideal conditions, iterative updates in the two parameter spaces (standardized stellar-parameter space and basis coefficient space) can be related through \refformula{PISP-CurveFit-PCA-2}, yielding nearly identical stellar-parameter inference results when numerical errors are negligible.
		\item \textbf{For non-rotation-invariant optimizers, PCA-Non-L1 ($n{=}25$) outperforms or matches the baseline strategy.}
		In the equal-dimensional, unregularized PCA-Non-L1 ($n{=}25$) case, non-rotation-invariant optimizers yield lower or comparable $\sigma(\Delta)$ relative to the baseline strategy. The underlying reason is that for these optimizers, the optimization paths corresponding to \refformula{PISP-CurveFit-PCA-loss1} and \refformula{PISP-CurveFit-PCA-loss2} can no longer be kept consistent through \refformula{PISP-CurveFit-PCA-2}. In this context, the decorrelation introduced by orthogonal reparameterization serves as a practical and effective data preprocessing technique. Similar ideas appear in regression analysis; for instance, \citet{Liang2022} demonstrated that decorrelation processing can improve parameter prediction accuracy while maintaining dimensionality. As a specific example, for Adam under the baseline strategy, introducing ReduceLROnPlateau does not reduce $\sigma(\Delta)$: comparing the RMSE of flux residuals at the optimal solution, we find that Adam with ReduceLROnPlateau reduces RMSE by only $\sim 2\times10^{-5}$ relative to Adam without ReduceLROnPlateau, already below the intrinsic RMSE limit of the spectral emulator (see \reffig{ANN_ResNet_flux_reproduced}). When optimization approaches this limit, further RMSE reduction does not translate to lower $\sigma(\Delta)$. Moreover, even with ReduceLROnPlateau, the baseline strategy yields higher $\sigma(\Delta)$ than PCA-Non-L1 under the same optimizer.
		
		\item \textbf{PCA-L1 consistently outperforms across all optimizers.}
		Across all tested optimizers, PCA-L1 achieves the lowest $\sigma(\Delta)$. This reduction is primarily driven by regularization constraints---selecting the regularization strength $\alpha$ in a data-driven manner on the validation or reference set to match data distribution and noise levels. Thus, the PCA-L1 strategy helps reduce $\sigma(\Delta)$ across different optimizer configurations.
	\end{itemize}
	
	\begin{figure*}[htp!]
		\centering
		\includegraphics[width=\textwidth,height=0.33\textheight, keepaspectratio]{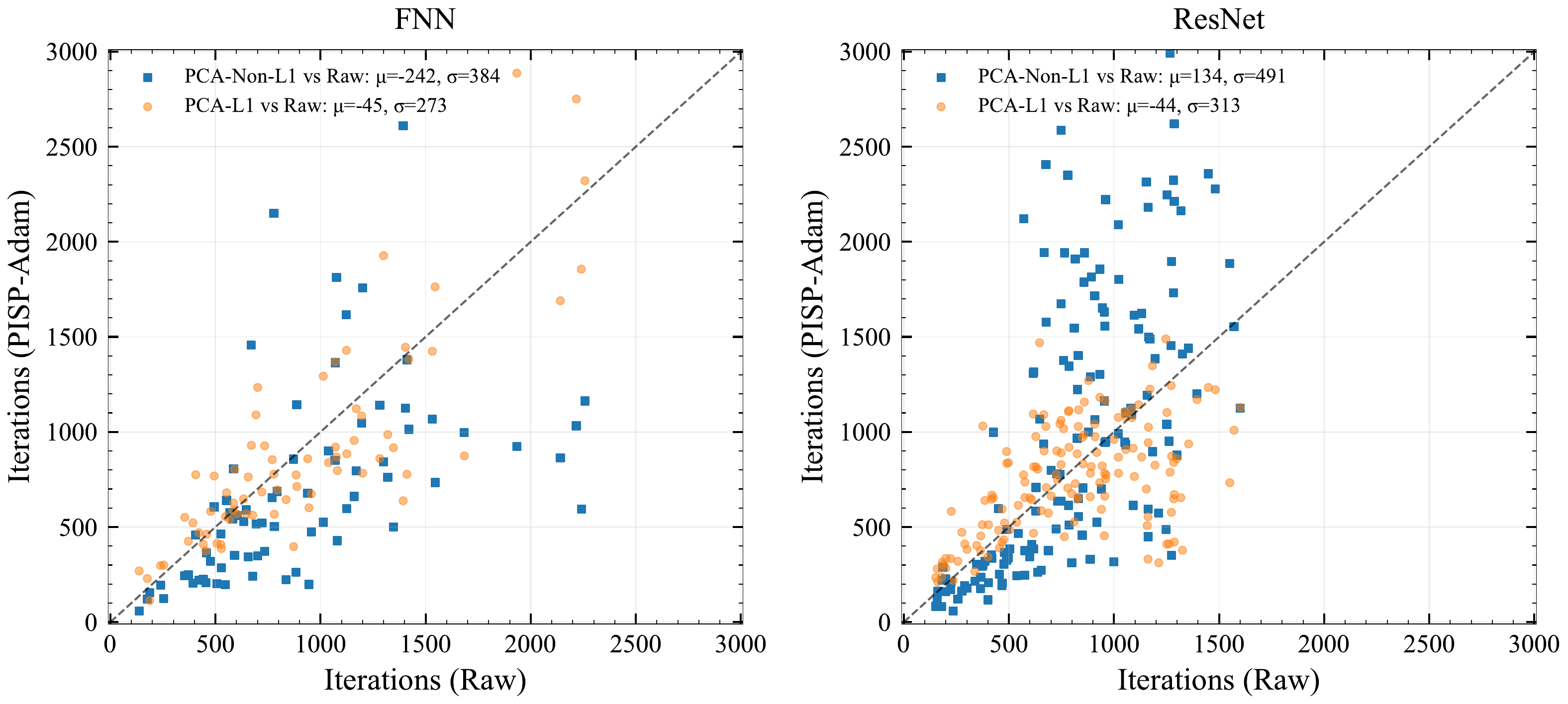}
		\caption{Comparison of iteration counts between PISP-Adam and  the baseline strategy for the inference of $25$-dimensional stellar parameters from $722{,}896$ APOGEE DR$17$ spectra. The left and right panels correspond to results using FNN and ResNet spectral emulators, respectively. The left panel includes $73$ groups ($72$ with $N{=}10{,}000$ spectra and one with $N{=}2896$), while the right panel includes $145$ groups ($144$ with $N{=}5000$ and one with $N{=}2896$). In each panel, the $\mu$ and $\sigma$ of differences between the iteration counts of PISP-Adam and those of Adam in the baseline strategy are computed using \texttt{astropy.stats.sigma\_clipped\_stats} (\texttt{sigma=3}, \texttt{maxiters=3}). The x-axis indicates the number of iterations required by Adam in the baseline strategy and the y-axis shows the iterations required by PISP-Adam in the principal-component space. For PCA-Non-L1, $n{=}7$; for PCA-L1, $\alpha{=}0.06$.}
		\label{PISP_Adam_iterations}
	\end{figure*}
	
	Based on these comparisons, under equal-dimensional orthogonal reparameterization on theoretical spectra, PISP's performance trends relative to the baseline strategy are generally consistent across optimizers, indicating that its improvements are primarily driven by reparameterization and regularization configurations during the inference stage rather than dependence on any specific optimizer setting.
	
	\subsection{Efficiency of Stellar Parameter Inference with PISP} \label{PISP-efficiency}
	To quantify the computational efficiency of PISP for large-scale parameter inference, we evaluated PISP-CurveFit and PISP-Adam on $722{,}896$ observed spectra and compared them with a baseline strategy. All experiments adopted the default \texttt{n\_jobs}, $N$, as well as the PCA projection bases and optimal hyperparameters determined in \refsubsection{pisp-hyperopt}. PISP-CurveFit was run on a blade server node with dual Intel Xeon Gold 5416S processors ($32$ cores, $64$ threads), and PISP-Adam was run on a server equipped with an NVIDIA GeForce RTX 3090. The same Python virtual environment and dependency versions were used across runs to ensure a fair comparison.
	
	As shown in \reftable{APOGEE_Time}, inference efficiency varies substantially across spectral emulator--optimizer--strategy combinations:
	\begin{itemize}
		\item \textbf{Spectral emulator efficiency.}
  		Owing to its lower computational complexity for spectral generation, FNN attains higher throughput than ResNet under the same optimizer and strategy. With the Adam optimizer, this gap is particularly pronounced, and FNN achieves a speedup exceeding $16\times$.
		\item \textbf{Optimizer efficiency.}
  		Adam supports GPU-accelerated parallel inference and, with an appropriate choice of $N$ (see \cite{Liang2026}, Figure~1), generally outperforms the CPU-based multithreaded CurveFit (the ResNet emulator with the PCA-Non-L1 strategy is the exception\footnote{For the ResNet emulator, PISP-CurveFit achieves an efficiency improvement of about a factor of $4$ relative to the baseline strategy. In contrast, PISP-Adam requires more iterations to converge (see \reffig{PISP_Adam_iterations}, right panel), leading to slightly lower efficiency. This may be related to the convergence behavior of Adam under the ResNet emulator, while a detailed analysis is beyond the scope of this work.}). Using the FNN emulator as an example, Adam yields more than $10\times$ speedup across all three strategies, reaching $\sim$$46\times$ under the PCA-L1 strategy ($0.49$ versus $22.45$ hr).
		\item \textbf{Strategy efficiency.}
  		Relative to the baseline strategy, PCA-Non-L1 reduces the optimization dimensionality from $25$ to $7$, thereby substantially reducing elapsed runtime (the ResNet + PCA-Non-L1 + Adam combination is an exception), with improvements as high as $\sim$$4\times$ ($20.78$ versus $4.99$ hr). In principal-component space, Adam also requires a substantially smaller number of iterations to reach convergence (see \reffig{PISP_Adam_iterations}); for the FNN emulator, PCA-Non-L1 reduces the mean iteration count by about $242$ relative to the baseline strategy. By contrast, the runtime under PCA-L1 changes only marginally, primarily because the decision-variable dimension is not reduced a priori and the regularizer introduces additional overhead.
	\end{itemize}

	In summary, without sacrificing accuracy, PISP delivers consistent efficiency gains across different emulator--optimizer configurations and is therefore suitable for high-throughput parameter inference on large spectroscopic datasets.
		
	\subsection{Future Improvements} \label{Future improvements}
	To further enhance the inference accuracy of high-dimensional stellar parameters and the adaptability to observed data, several avenues remain for improving PISP. From the perspectives of parameter-space modeling, adaptation to the observational domain, result reliability, and cross-survey evaluation, we propose the following directions:
	\begin{itemize}
		\item \textbf{Nonlinear embeddings for parameter-space representation.}
		The current PISP employs linear projection to mitigate the impact of variable correlations on parameter inference. However, when nonlinear dependencies exist among elemental abundances, its expressiveness is limited. Building on successful applications of nonlinear embeddings in optimization problems \citep{Tripp2020,Antonov2022,Maus2022,Chu2024,Boyar2024}, future work can explore kernel principal component analysis and variational autoencoders within the PISP framework to strengthen the characterization of nonlinear structure and further improve inference accuracy for certain parameter dimensions.
		\item \textbf{Observational-domain adaptation and robustness.}
		Due to distributional differences between synthetic and observed spectra, PISP may underperform on observational data relative to its performance on synthetic test sets. To improve robustness in observations, one may incorporate the \emph{Clean} strategy from the Python-based LAMOST Stellar Parameter Pipeline (PyLASP; \citealt{Liang2026}), which adaptively identifies and removes anomalous pixels with large flux differences during inference, retaining only high-consistency regions for fitting and thereby enhancing noise robustness. In parallel, the Cycle-StarNet method \citep{O’Briain2021} can be leveraged to construct an explicit cross-domain mapping between synthetic and observed spectra, correcting systematic differences at the model level and further improving the inference accuracy of strategies such as PISP on real observations.
		\item \textbf{Parameter recommendation and per-label quality flags.}
		Although PISP has produced high-dimensional parameter inferences on large-scale observed spectra, a systematic recommendation mechanism is not yet in place. In practice, some spectra may have low S/Ns or certain parameters may deviate in accuracy or physical consistency due to inherent limitations of both the synthetic spectral models and the inference methodology. To improve the scientific usability of the catalog, we recommend introducing a per-parameter recommendation flag (\texttt{FLAG}) that encodes a credibility level. This mechanism can integrate factors such as spectral quality, objective-function convergence, parameter uncertainty, and distributional plausibility, thereby providing a more robust catalog for subsequent studies. Detailed design and implementation are left for future work.
		\item \textbf{End-to-end evaluation of official pipelines for high-dimensional stellar-parameter inference.}
		PISP has not yet undergone systematic evaluation within the official pipelines of large spectroscopic surveys for high-dimensional stellar-parameter inference. Surveys differ substantially in preprocessing, model interfaces, and runtime environments, while the availability and stability of pipelines and training or reference resources also vary. Even where some components are open source, additional engineering integration and recalibration would be required, which is beyond the scope of this study. To ensure fairness and reproducibility, we therefore benchmark PISP against multiple baseline strategies under strictly controlled conditions---using identical data, spectral emulators, and optimization settings. Once interfaces stabilize and integration becomes feasible, we will extend the evaluation to end-to-end validation within native survey pipelines and release the corresponding scripts and results.
	\end{itemize}
	
	\section{Conclusions} \label{Conclusions}
	This work proposes PISP, a projection-space-assisted framework for high-dimensional stellar-parameter inference. By mapping standardized stellar parameters into a projection space via linear transformation, PISP improves both inference accuracy and efficiency. To accommodate different computing environments and data scales, we implement PISP-CurveFit for fast per-object inference and PISP-Adam for large-scale parallel computation. Using Kurucz synthetic spectra and APOGEE DR$17$ observed spectra, together with FNN and ResNet spectral emulators, we conduct a systematic evaluation of PISP. The main conclusions are as follows:
	\begin{itemize}
		\item \textbf{Dataset-specific hyperparameter tuning is necessary for optimal performance.}
		PISP is sensitive to the number of basis vectors $n$, the L1 regularization strength $\alpha$, the learning rate $lr$, and the convergence threshold $\epsilon$. On Kurucz synthetic spectra, a larger $n$ and a smaller $\alpha$ help preserve beneficial directions; on APOGEE observations, a larger $\alpha$ and a smaller $n$ are preferred. Therefore, $n$ and $\alpha$ should be tuned independently according to the task and data distribution. In addition, $lr$ and $\epsilon$ primarily affect convergence speed and stability. In this work, we adopt $lr{=}0.01$ and $\epsilon{=}10^{-8}$ uniformly across datasets.
		\item \textbf{PISP delivers concurrent gains in accuracy and computational efficiency.}
		On the Kurucz test set, PCA-L1 performs best, reducing $\sigma(\Delta)$ by at least $0.01$ dex for $12$ of $20$ elemental abundances across all emulator-optimizer combinations, with [N/H], [O/H], [Na/H], [Co/H], [P/H], [V/H], [Cu/H] showing reductions of $0.05{-}0.72$ dex. On APOGEE DR$17$ observations, PCA-Non-L1 reduces $\sigma(\Delta)$ by more than $30$ K for $T_{\mathrm{eff}}$ and by at least $0.01$ dex for $9$ of $17$ elemental abundances across all emulator-optimizer combinations, with [O/H], [Na/H], [V/H] showing reductions of $0.05{-}0.20$ dex, exhibiting higher agreement with the official catalog and confirming PISP's applicability to real survey data. For computational efficiency, PCA-Non-L1 achieves up to $\sim$$4\times$ acceleration for large-scale inference on APOGEE data, while PISP-Adam under the PCA-L1 strategy accelerates by up to $46\times$ relative to PISP-CurveFit. These results indicate that PISP balances accuracy and efficiency across datasets and emulator-optimizer combinations, providing a practical solution for high-dimensional stellar-parameter inference in large spectroscopic surveys.
		\item \textbf{The regularization choice dictates basis vectors selection and guides strategy preference.}
		L1 and Non-L1 adopt different selection mechanisms: the former induces sparsity in coefficients across all $25$ basis vectors via L1 regularization and can dynamically adjust the effective dimensionality for each individual spectrum; the latter fixes a subset of high-variance principal components or active-subspace directions, assuming by default that dominant directions have consistent importance across all samples. Experiments show that PCA-L1 achieves lower $\sigma(\Delta)$ when the training and test distributions are aligned; under distribution shift in observed data, PCA-Non-L1 achieves slightly lower $\sigma(\Delta)$ than PCA-L1, with lower computational cost.
		\item \textbf{PISP integrates seamlessly across emulators and optimizers without retraining.}
		Whether using PISP-CurveFit or PISP-Adam, and either the FNN or ResNet spectral emulator, PISP reduces $\sigma(\Delta)$ relative to the baseline strategy without retraining the spectral emulator and thus can be readily integrated into different parameter-inference pipelines.
	\end{itemize}
	
	In summary, PISP demonstrates advantages in accuracy, efficiency, and cross-domain adaptability, providing a general and efficient optimization framework for high-dimensional stellar-parameter inference. In future work, PISP can be integrated with PyLASP and applied to high-dimensional stellar-parameter inference for LAMOST spectra.
	
	\section*{Acknowledgments}
	We thank Yuan-Sen Ting for providing the Kurucz synthetic spectra, without which this work would not have been possible.
	
	This work is supported by the National Natural Science Foundation of China (12273078, 12273075, and 12411530071), the National Key Research and Development Program (2025YFF0510602), and the National Astronomical Observatories of the Chinese Academy of Sciences (No. E4ZR0516). We also acknowledge support from a Royal Society IEC\textbackslash NSFC\textbackslash233140 exchange grant.
	
	Funding for the Sloan Digital Sky 
	Survey IV has been provided by the 
	Alfred P. Sloan Foundation, the U.S. 
	Department of Energy Office of 
	Science, and the Participating 
	Institutions. 
	
	SDSS-IV acknowledges support and 
	resources from the Center for High 
	Performance Computing  at the 
	University of Utah. The SDSS 
	website is www.sdss4.org.
	
	SDSS-IV is managed by the 
	Astrophysical Research Consortium 
	for the Participating Institutions 
	of the SDSS Collaboration including 
	the Brazilian Participation Group, 
	the Carnegie Institution for Science, 
	Carnegie Mellon University, Center for 
	Astrophysics | Harvard \& 
	Smithsonian, the Chilean Participation 
	Group, the French Participation Group, 
	Instituto de Astrof\'isica de 
	Canarias, The Johns Hopkins 
	University, Kavli Institute for the 
	Physics and Mathematics of the 
	Universe (IPMU) / University of 
	Tokyo, the Korean Participation Group, 
	Lawrence Berkeley National Laboratory, 
	Leibniz Institut f\"ur Astrophysik 
	Potsdam (AIP),  Max-Planck-Institut 
	f\"ur Astronomie (MPIA Heidelberg), 
	Max-Planck-Institut f\"ur 
	Astrophysik (MPA Garching), 
	Max-Planck-Institut f\"ur 
	Extraterrestrische Physik (MPE), 
	National Astronomical Observatories of 
	China, New Mexico State University, 
	New York University, University of 
	Notre Dame, Observat\'ario 
	Nacional / MCTI, The Ohio State 
	University, Pennsylvania State 
	University, Shanghai 
	Astronomical Observatory, United 
	Kingdom Participation Group, 
	Universidad Nacional Aut\'onoma 
	de M\'exico, University of Arizona, 
	University of Colorado Boulder, 
	University of Oxford, University of 
	Portsmouth, University of Utah, 
	University of Virginia, University 
	of Washington, University of 
	Wisconsin, Vanderbilt University, 
	and Yale University.
	
	\bibliographystyle{aasjournal}  
	\footnotesize
	\bibliography{refs}
\end{document}